\pdfoutput=1

\documentclass[11pt,twoside,a4paper,cmspaper,final,collab]{cms-tdr}
\def\svnVersion{4265e4c}\def\svnDate{2020/03/11}\def\cmsCernNoTag{CERN-EP-2019-231}\def\cmsCernDate{\today}\def\cmsMessage{"Published in the Journal of High Energy Physics as \href{http://dx.doi.org/10.1007/JHEP03(2020)065}{\doi{10.1007/JHEP03(2020)065}.}"}

\begin{document}\cmsNoteHeader{HIG-18-023}

\hyphenation{had-ron-i-za-tion}
\hyphenation{cal-or-i-me-ter}
\hyphenation{de-vices}
\ifthenelse{\boolean{cms@external}}{\providecommand{\cmsRight}{bottom\xspace}}{\providecommand{\cmsRight}{right\xspace}}
\providecommand{\NA}{\ensuremath\text{{ --- }}\xspace}
\providecommand{\CL}{CL\xspace}
\providecommand{\cmsTable}[1]{\resizebox{\textwidth}{!}{#1}}
\newlength\cmsTabSkip\setlength{\cmsTabSkip}{1ex}
\newcommand{\mAvis}{\ensuremath{m^\text{vis}_{\ell\ell\PGt\PGt}}\xspace}
\newcommand{\mA}{\ensuremath{m_\text{A}}\xspace}
\newcommand{\PA}{\ensuremath\text{{A}}\xspace}
\newcommand{\mtt}{\ensuremath{m_{\PGt\PGt}}\xspace}
\newcommand{\mlltt}{\ensuremath{m_{\ell\ell\PGt\PGt}}\xspace}
\newcommand{\LTH}{\ensuremath{L_\text{{T}}^\text{{h}}}\xspace}
\newcommand{\mH}{\ensuremath{m_{\Ph}}\xspace}
\pretolerance=10000

\cmsNoteHeader{HIG-18-023}
\title{Search for a heavy pseudoscalar Higgs boson decaying into a 125\GeV Higgs boson and a \PZ boson in final states with two tau and two light leptons at \texorpdfstring{$\sqrt{s}=13\TeV$}{sqrt(s) = 13 TeV}}

\date{\today}

\abstract{
A search is performed for a pseudoscalar Higgs boson, $\PA$, decaying into a 125\GeV Higgs boson \Ph and a \PZ boson.
The \Ph boson is specifically targeted in its decay into a pair of tau leptons, while the \PZ boson decays into a pair of electrons or muons.
A data sample of proton-proton collisions collected by the CMS experiment at the LHC at $\sqrt{s} = 13\TeV$  is used, corresponding to an integrated luminosity of 35.9\fbinv.
No excess above the standard model background expectations is observed in data.
A model-independent upper limit is set on the product of the gluon fusion production cross section for the $\PA$ boson and
the branching fraction to $\PZ\Ph\to\ell\ell\PGt\PGt$.
The observed upper limit at 95\% confidence level ranges from 27 to 5\unit{fb} for $\PA$ boson masses from 220 to 400\GeV, respectively.
The results are used to constrain the extended Higgs sector parameters for two benchmark scenarios of the minimal supersymmetric standard model.
}

\hypersetup{
pdfauthor={CMS Collaboration},
pdftitle={Search for a heavy pseudoscalar Higgs boson decaying into a 125 GeV Higgs boson and a Z boson in final states with two tau and two light leptons at sqrt(s) =13 TeV},
pdfsubject={CMS},
pdfkeywords={CMS, physics, Higgs, Higgs associated production}}

\maketitle

\section{Introduction\label{sec:intro}}
In the standard model (SM)~\cite{Glashow,SM1,SM3}, the Brout--Englert--Higgs
mechanism~\cite{Englert:1964et,Higgs:1964ia,Higgs:1964pj,Guralnik:1964eu,Higgs:1966ev,Kibble:1967sv}
is responsible for the electroweak symmetry breaking, and it predicts the
existence of the Higgs boson. A Higgs boson with a mass around 125\GeV was discovered
by the ATLAS and CMS Collaborations in 2012~\cite{ATLASobservation125,CMSobservation125,CMSobservation125Long}.
The best measurement of the Higgs boson mass to date, $125.26\pm0.21\GeV$, comes from a partial Run 2 data set analysis by the CMS Collaboration~\cite{Sirunyan:2017exp};
the result is consistent with the earlier Run 1 combined measurement by the ATLAS and CMS Collaborations~\cite{Aad:2015zhl} and the recent results by the ATLAS Collaboration~\cite{Aaboud:2018wps}.
The couplings of the observed boson have been studied extensively, and are found to be compatible with the SM expectation~\cite{Aad:2019mbh, Sirunyan:2018koj}.

The observation of the Higgs boson has not only given closure to the search for particles described by the SM,
but also constrains the beyond-the-SM theories proposed to explain some of the open questions in particle physics.
A class of simple extensions of the SM, two-Higgs-doublet models (2HDMs), predicts the existence of five Higgs bosons~\cite{Lee:1973iz, Branco:2011iw}.
Two of these five particles are CP-even Higgs bosons (\Ph and \PH), and thus either of them could correspond to the observed particle.
The properties of the observed state can be used to exclude regions of the parameter space of 2HDMs. Further constraints can be placed by performing searches
for the four additional Higgs bosons, namely the scalar \PH, the CP-odd Higgs boson $\PA$, and two charged Higgs bosons
$\PH^{\pm}$. Moreover, 2HDMs are a prerequisite for the minimal supersymmetric standard model (MSSM) where the extended Higgs sector at tree-level is fully defined by two parameters, conventionally
chosen to be the ratio of the vacuum expectation values of the two Higgs doublets (\tanb) and the mass of the pseudoscalar $\PA$ ($\mA$).

In the MSSM, given that the mass of the \Ph boson is as large as 125\GeV, the scale of the soft supersymmetry breaking masses can be larger than 1\TeV.
This is a reasonable assumption based on the nonobservation of supersymmetric particles at the CERN LHC thus far.
In many of the MSSM benchmark scenarios typically studied, the predicted mass of the Higgs boson is lower than 125\GeV in the low $\tanb$ region~\cite{Bahl:2019ago}.
We study two MSSM benchmark scenarios that can accommodate these constraints in most of the $\mA$--$\tanb$ plane: $\mathrm{M^{125}_{\Ph,EFT}}$~\cite{Bahl:2019ago} and hMSSM~\cite{Djouadi:2013vqa, Maiani:2013hud, Djouadi:2013uqa, Djouadi:2015jea}.
The Higgs sector predictions of the $\mathrm{M^{125}_{\Ph,EFT}}$ scenario are derived from a 2HDM effective field theory framework,
with a supersymmetric mass scale that can reach up to $10^{16}\GeV$, in order for the Higgs boson mass to be compatible with 125\GeV in the low $\tanb$ region.
In the hMSSM scenario, by requiring $\mH=125\GeV$, the dominant radiative corrections to the Higgs boson mass become fixed, which are then used to determine the masses and couplings of the other Higgs bosons.

\begin{figure}[h!]
\centering
  \includegraphics[width=0.34\textwidth]{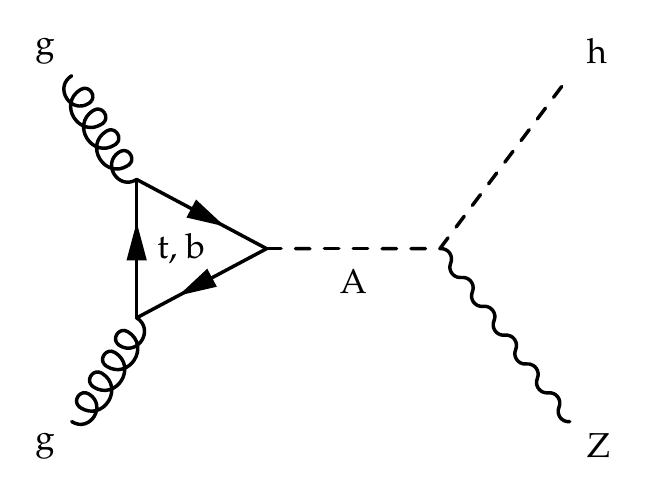}
  \includegraphics[width=0.34\textwidth]{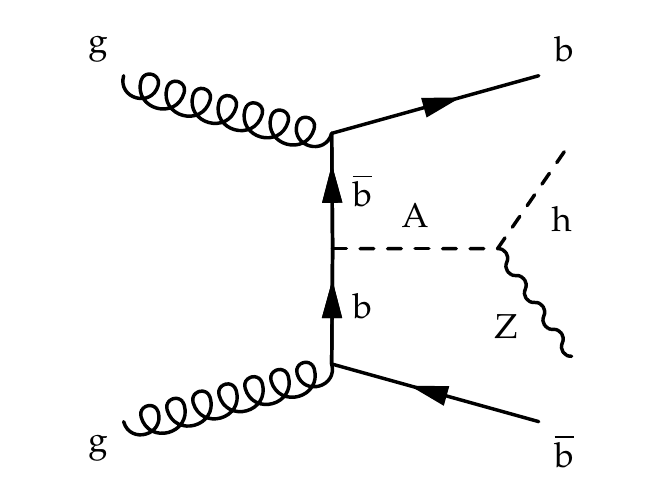}
 \caption{Feynman diagrams for two dominant production processes for the pseudoscalar $\PA$ boson: gluon fusion (left) and associated production with $\cPqb$ quarks (right).
In both cases the $\PA$ boson decays into a 125\GeV Higgs boson and a \PZ boson.
 }
 \label{fig:productionAboson}
\end{figure}

The parameter spaces of these benchmark scenarios can be explored by studying processes producing an experimentally accessible signature with a 125\GeV Higgs boson.
One such process is the decay of the $\PA$ boson into a 125\GeV Higgs boson and a \PZ boson.
In the parameter space region with low \tanb values, this decay has a substantial branching fraction. For $\tanb\lesssim5$ the $\PA$ boson is produced mainly in gluon fusion ($\Pg\Pg\to\PA$),
but for higher \tanb values the associated production with $\cPqb$ quarks ($\bbbar\PA$) becomes dominant. The Feynman diagrams for both production processes are shown in Fig.~\ref{fig:productionAboson}.

This paper reports on a search for the pseudoscalar $\PA$ boson decaying into a 125\GeV Higgs boson \Ph and a \PZ boson in proton-proton ($\Pp\Pp$)
collisions at $\sqrt{s}=13\TeV$. The search is based on a data set collected in 2016 by the CMS experiment, corresponding to an integrated luminosity of 35.9\fbinv.
The analysis is primarily sensitive to the assumed gluon fusion production of the $\PA$ boson, but the associated production with $\cPqb$ quarks is included in the interpretation of the results.
The studied signal mass range begins at 220\GeV because the $\PA$ boson must be massive enough to decay into the considered $\PZ\Ph$ state.
The mass range extends up to 400\GeV, slightly above where the mass of the $\PA$ boson exceeds twice the top quark mass. In this region the $\PA\to\ttbar$ decay channel is expected to dominate.

Previous searches for the $\PA\to\PZ\Ph$ process, performed by the ATLAS and CMS Collaborations, considered a final state with two tau leptons from the \Ph boson decay,
and used data sets collected in $\Pp\Pp$ collisions at $\sqrt{s}=8$\TeV~\cite{Aad:2015wra, Khachatryan:2015tha}.
The ATLAS and CMS Collaborations have also searched for the pseudoscalar $\PA$ decaying into the same intermediate
$\PZ\Ph$ state but with the Higgs boson \Ph decaying into a pair of bottom quarks in $\Pp\Pp$ collisions at $\sqrt{s}=13\TeV$~\cite{Aaboud:2017cxo, Sirunyan:2019xls}.
These analyses set both model-independent and model-dependent limits in the context of 2HDMs.

In this search, the Higgs boson is sought in its decay into a pair of tau leptons.
Four possible $\PGt\PGt$ decay channels of the Higgs boson are considered: $\Pe\tauh$, $\Pgm\tauh$, $\tauh\tauh$, and
$\Pe\Pgm$, where \tauh denotes a tau lepton decaying hadronically. Throughout the paper, neutrinos are omitted from the notation of the final states.
These four decay channels are combined with the \PZ boson decays into two light
leptons, \ie, $\PZ\to\ell^+\ell^-$ ($\ell=\Pe,\Pgm$),
resulting in eight distinct final states of the $\PA$ boson decay.
To account for the missing transverse momentum that results from the neutrinos in the final states, we
use the \textsc{svfit} algorithm~\cite{Bianchini:2016yrt} to reconstruct the four-vector of the Higgs boson while constraining its mass to 125\GeV.
Compared to the previous result presented by the CMS Collaboration~\cite{Khachatryan:2015tha},
this novel approach significantly increases the sensitivity of the search.

\section{The CMS detector}
The central feature of the CMS apparatus is a superconducting solenoid of 6\unit{m} internal diameter,
providing a magnetic field of 3.8\unit{T}. Within the solenoid volume are a silicon pixel and strip tracker,
a lead tungstate crystal electromagnetic calorimeter (ECAL), and a brass and scintillator hadron calorimeter,
each composed of a barrel and two endcap sections.
Forward calorimeters extend the pseudorapidity ($\eta$) coverage provided by the barrel and endcap detectors.
Muons are detected in gas-ionization chambers embedded in the steel flux-return yoke outside the solenoid.
Events of interest are selected using a two-tiered trigger
system~\cite{Khachatryan:2016bia}.  The first level (L1), composed of custom hardware processors, uses information from the calorimeters
and muon detectors to select events at a rate of around 100\unit{kHz} within a time interval of less than $4\mus$.
The second level, known as the high-level trigger (HLT), consists of a farm of processors running a version of the full event reconstruction
software optimized for fast processing, and reduces the event rate to around 1\unit{kHz} before data storage.
A more detailed description of the CMS detector, together with a definition of
the coordinate system used and the relevant kinematic variables, can be found
in Ref.~\cite{Chatrchyan:2008zzk}.

\section{Simulated samples and models}
Simulated signal events with a pseudoscalar Higgs boson $\PA$ produced in gluon fusion ($\Pg\Pg\to\PA$),
decaying into a 125\GeV Higgs boson and a \PZ
boson and finally into two tau and two leptons (electrons or muons) are generated at leading order
(LO) using \MGvATNLO v2.4.2~\cite{Alwall2014}. The considered $\PA$ boson mass points are within 220--400\GeV, as
in this mass range the $\PA\to\PZ\Ph$ decay becomes predominant.
The samples are based on the $m_{\Ph}^{\mathrm{mod+}}$ model~\cite{Carena:2013ytb}, assuming a low value of $\tanb$ ($\sim$2).
The generated width of the $\PA$ boson is small compared to the instrumental resolution for all masses.
Additional signal events are simulated for a 300\GeV $\PA$ boson produced in association with $\cPqb$ quarks ($\bbbar\PA$)
and are used only to study the selection efficiency,
necessary for setting model-dependent limits, as explained in Section~\ref{sec:results}.

The background processes consist of all SM processes with nonnegligible yield in the
studied final states, including the Higgs boson production through processes predicted in the SM (e.g.\ $\PZ\Ph$, $\PW\Ph$, $\ttbar\Ph$).
The background processes with a Higgs boson decaying into two tau leptons, produced in association with a \PW or
\PZ boson ($\PW\Ph$ or $\PZ\Ph$), are generated at next-to-LO (NLO) in perturbative quantum chromodynamics (QCD) with the \textsc{POWHEG}
2.0~\cite{Nason:2004rx,Frixione:2007vw, Alioli:2010xd, Alioli:2010xa, Alioli:2008tz}
generator extended with the MiNLO procedure~\cite{Luisoni:2013kna}.
The contribution from Higgs boson events produced via gluon fusion or vector boson fusion and decaying into two tau leptons is negligible.
The transverse momentum ($\pt$) distribution of the Higgs boson in the \textsc{POWHEG} simulations is tuned to match closely
the next-to-NLO (NNLO) plus next-to-next-to-leading-logarithmic prediction in the
\textsc{HRes 2.3} generator~\cite{deFlorian:2012mx,Grazzini:2013mca}.
The production cross sections and branching fractions for the SM Higgs
boson production and their corresponding uncertainties are taken from
Refs.~\cite{deFlorian:2016spz,Denner:2011mq,Ball:2011mu}.

The background samples for $\ttbar\Ph$, $\ttbar$, $\PW\PZ$, and $\cPq\cPq\to\PZ\PZ$, as well as $\PW\Ph\to\PW\PW\PW$, $\PW\Ph\to\PW\PZ\PZ$,
$\PZ\Ph\to\PZ\PW\PW$, $\PZ\Ph\to\PZ\PZ\PZ$, and $\Pg\Pg\to\Ph\to\PZ\PZ$ processes,
are generated at NLO with \textsc{POWHEG} 2.0.
The $\Pg\Pg\to\PZ\PZ$ process is generated at LO with
\textsc{mcfm}~\cite{Campbell:2010ff}. Samples for the $\cPq\cPq\to\PZ\PZ$ and $\Pg\Pg\to\PZ\PZ$ processes
include all SM events with two \PZ bosons in the final states except the ones from the $\Pg\Pg\to\Ph\to\PZ\PZ$ process.
 The \MGvATNLO 2.2.2 or 2.3.3 generator is used for
triboson, $\PZ+\text{jets}$, $\ttbar\PW$, and $\ttbar\PZ$ production,
with the jet matching and merging scheme applied either at NLO with the FxFx algorithm~\cite{Frederix:2012ps}
or at LO with the MLM algorithm~\cite{Alwall:2007fs}.
The generators are interfaced with \PYTHIA 8.212~\cite{Sjostrand:2014zea}
to model the parton showering and fragmentation, as well as the decay of the tau leptons.
The \PYTHIA parameters affecting the description of the underlying event are
set to the {CUETP8M1} tune~\cite{Khachatryan:2015pea}.
The set of parton distribution functions (PDFs) used in the simulation is
NNPDF3.0~\cite{Ball:2011uy}.

The generated events are processed through a simulation of the CMS detector based on
\ \GEANTfour~\cite{Agostinelli:2002hh}, and are reconstructed with the same algorithms
that are used for data.
The simulated samples include additional $\Pp\Pp$ interactions per bunch
crossing, referred to as in-time pileup. The effect of inelastic collisions happening in the preceding
and subsequent bunch crossings (out-of-time pileup) is also considered.
The effect of pileup is taken into account by generating concurrent minimum bias
collision events. The simulated events are weighted such that the distribution of the number
of pileup interactions matches with that observed in data. The pileup distribution is
estimated from the measured instantaneous luminosity for each bunch crossing, resulting in
an average of approximately 23 interactions per bunch crossing.

To produce model-dependent interpretations of the results described in Section~\ref{sec:results},
we utilize production cross section and branching fraction calculations for the pseudoscalar $\PA$ in the $\mathrm{M^{125}_{\Ph,EFT}}$ and hMSSM scenarios.
In the $\mathrm{M^{125}_{\Ph,EFT}}$ scenario, Higgs boson masses and mixing parameters (and effective Yukawa couplings) were calculated with a yet unpublished version of \textsc{FeynHiggs}
based on version 2.14.3~\cite{Heinemeyer:1998yj, Heinemeyer:1998np, Degrassi:2002fi, Frank:2006yh, Hahn:2013ria, Bahl:2019ago}.

For the gluon-gluon fusion process in the $\mathrm{M^{125}_{\Ph,EFT}}$ (hMSSM) scenario, inclusive cross sections are obtained with \textsc{SusHi} 1.7.0 (1.4.1)~\cite{Harlander:2012pb, Harlander:2016hcx}, which
includes supersymmetric NLO QCD corrections~\cite{Spira:1995rr, Harlander:2004tp, Harlander:2005rq, Degrassi:2010eu, Degrassi:2011vq, Degrassi:2012vt},
NNLO QCD corrections for the top quark contribution in an effective theory of a heavy top quark~\cite{Harlander:2002wh,Anastasiou:2002yz,Ravindran:2003um,Harlander:2002vv,Anastasiou:2002wq} and
electroweak effects from light quarks~\cite{Aglietti:2004nj, Bonciani:2010ms}.

Inclusive $\bbbar\PA$ production cross sections at NNLO QCD accuracy in the five-flavor scheme are calculated with \textsc{SusHi}, based on \textsc{bbh@nlo}~\cite{Harlander:2003ai}.
The results are combined with the $\bbbar\PA$ cross section calculation at NLO in QCD in the four-flavor scheme~\cite{Dittmaier:2003ej, Dawson:2003kb}
using the Santander matching scheme~\cite{Harlander:2011aa} for the hMSSM scenario, and matched predictions~\cite{Bonvini:2015pxa, Bonvini:2016fgf, Forte:2015hba, Forte:2016sja} for the $\mathrm{M^{125}_{\Ph,EFT}}$ scenario.

In the hMSSM scenario, branching fractions are solely computed with \textsc{HDECAY} 6.40~\cite{Djouadi:1997yw, Djouadi:2006bz, Djouadi:2018xqq},
whereas the $\mathrm{M^{125}_{\Ph,EFT}}$ scenario relies on a yet unpublished version of \textsc{FeynHiggs} based on version 2.14.3~\cite{Heinemeyer:1998yj, Heinemeyer:1998np, Degrassi:2002fi, Frank:2006yh, Hahn:2013ria, Bahl:2019ago}.

\section{Event reconstruction}\label{sec:reconstruction}
Both observed and simulated events are reconstructed using the
particle-flow (PF) algorithm~\cite{Sirunyan:2017ulk}.
The particle-flow algorithm aims to reconstruct and identify each individual particle in an event, with an optimized combination of information from the various elements of the CMS detector.
In this process the reconstructed PF objects include photons, electrons, muons, neutral
hadrons, and charged hadrons.

Higher-level objects are reconstructed from
combinations of the PF objects. For example, jets are reconstructed with an anti-\kt clustering algorithm implemented
in the \FASTJET library~\cite{Cacciari:2011ma, Cacciari:fastjet2}.
The reconstruction is based on the clustering of PF objects with
a distance parameter of 0.4. Charged PF objects are required to
be associated with the primary vertex of the interaction.
The reconstructed vertex with the largest value of summed physics-object $\pt^2$ is taken to be the primary $\Pp\Pp$ interaction vertex.
The physics objects are the jets, clustered using the jet finding algorithm~\cite{Cacciari:2008gp,Cacciari:2011ma} with the tracks assigned to the vertex as inputs,
and the associated missing transverse momentum, taken as the negative vector sum of the \pt of those jets.
Jet energy corrections are derived from simulation studies so that the average measured response of jets becomes identical to that of particle level jets.
In situ measurements of the momentum balance in dijet, photon$+$jet, $\PZ+\text{jet}$, and multijet events are used to determine any residual differences between the jet
energy scale in data and in simulation, and appropriate corrections are applied~\cite{Khachatryan:2016kdb}.

While neutrinos cannot be detected directly, they contribute to the missing transverse momentum.
The missing transverse momentum vector $\ptvecmiss$ is computed as the negative
vector sum of the transverse momenta of all the PF objects in an event~\cite{Sirunyan:2019kia}.
The \ptvecmiss is modified to account for corrections to the energy scale of the reconstructed jets in the event.

Electrons are identified by a multivariate analysis (MVA) discriminant that requires as input several quantities describing the track quality,
the shapes of the energy deposits in the ECAL, and the compatibility of the measurements from the tracker and the ECAL~\cite{Khachatryan:2015hwa}.
Muon identification relies on the number of hits in the inner tracker and the muon systems,
and on the quality of the reconstructed tracks~\cite{Sirunyan:2018fpa}. Electrons and muons selected in this analysis
are required to be consistent with originating from the primary vertex.

A lepton isolation discriminant $I^{\ell}$ is defined to reject nonprompt or misidentified leptons ($\ell=\Pe,\Pgm$):
\begin{linenomath}
\begin{equation}
I^{\ell} \equiv \frac{\sum_\text{charged}  \PT + \max\left( 0, \sum_\text{neutral}  \PT
                 - \frac{1}{2} \sum_\text{charged, PU} \PT  \right )}{\PT^{\ell}},
\label{eq:reconstruction_isolation}
\end{equation}
\end{linenomath}
where $\PT^{\ell}$ stands for the $\pt$ of the lepton. The variable $\sum_\text{charged}  \PT$ is the scalar sum of the
transverse momenta of the charged particles originating from
the primary vertex and located in a cone of size
$\Delta R = \sqrt{\smash[b]{(\Delta \eta)^2 + (\Delta \phi)^2}} = 0.3$\,(0.4)
centered on the electron (muon) direction, where $\phi$ is the azimuthal angle in radians. The sum
$\sum_\text{neutral}  \PT$ represents
a similar quantity for neutral particles.
The scalar sum of the transverse
momenta of charged hadrons originating from pileup vertices in the cone,
$\sum_\text{charged, PU} \PT$, is used to estimate the contribution of photons and neutral hadrons originating from pileup vertices.
The factor of $1/2$ corresponds approximately to the ratio of neutral- to charged-hadron
production in the hadronization process
of inelastic $\Pp\Pp$ collisions.
The isolation requirements based on $I^{\ell}$ are described in the following section.

The combined secondary vertex algorithm~\cite{Sirunyan:2017ezt} is used to identify jets that are likely to have originated from a bottom quark
(``$\cPqb$-tagged jets''). In this algorithm, the secondary vertices associated with the jet and
the track-based lifetime information are given as inputs to an MVA discriminant designed
for $\cPqb$ jet identification. Differences in the $\cPqb$ tagging
efficiency between data and simulation are taken into account by applying a set of $\pt$-dependent
correction factors to the simulated events~\cite{Sirunyan:2017ezt}.
The identification efficiency for genuine $\cPqb$ jets in this analysis is approximately 63\%, whereas the misidentification probability 
for $\cPqc$ (light-flavor or gluon) jets is approximately 12 (0.9)\% for jet $\pt>20\GeV$.

Anti-\kt jets seed the hadron-plus-strips algorithm~\cite{Khachatryan:2015dfa, Sirunyan:2018pgf}
which is used to reconstruct \tauh candidates.
A hadronic decay of a tau lepton can result in one or more charged hadrons, and additional $\PGpz$ particles.
These $\PGpz$s are reconstructed by clustering electromagnetic deposits in the ECAL into ``strips`` in the $\eta$-$\phi$ plane.
The strips are elongated in the $\phi$ direction to contain the calorimeter signatures of converted photons from neutral
pion decays.
The algorithm reconstructs \tauh candidates based on the number of tracks and the number of strips
representing the number of charged hadrons (``prongs'') and the number of $\PGpz$s present in the decay.
The \tauh candidates used in this analysis are reconstructed in three decay modes: 1-prong, 1-prong+$\PGpz\mathrm{s}$, and 3-prong.

To suppress objects (jets and leptons) misidentified as \tauh candidates, an MVA
discriminant~\cite{Sirunyan:2018pgf} including calorimetric information, isolation sums, and lifetime information is used.
A misidentification rate for quark- and gluon-initiated jets of less than 1\% is achieved
within a \pt range typical of a \tauh candidate originating from an \Ph boson. At the same time, an efficiency for selecting \tauh candidates
of $\approx$70\% can be achieved for \tauh candidates passing the decay mode reconstruction discussed above.
To further suppress electrons and muons misidentified as \tauh candidates, dedicated criteria based on the consistency between the measurements in the
tracker, the calorimeters, and the muon detectors are applied~\cite{Khachatryan:2015dfa, Sirunyan:2018pgf}.
The \tauh energy scale is measured from $\cPZ\to\PGt\PGt$ events
and the correction is propagated to the simulation for each decay mode.
A ``tag-and-probe'' measurement~\cite{Sirunyan:2018pgf} in $\PZ\to\ell\ell$ events, where one of the $\ell$ is misidentified as a \tauh candidate,
is used to correct the energy scale of electrons and muons misidentified as \tauh candidates in simulation.

\begin{figure}[h!]
\centering
  \includegraphics[width=0.65\textwidth]{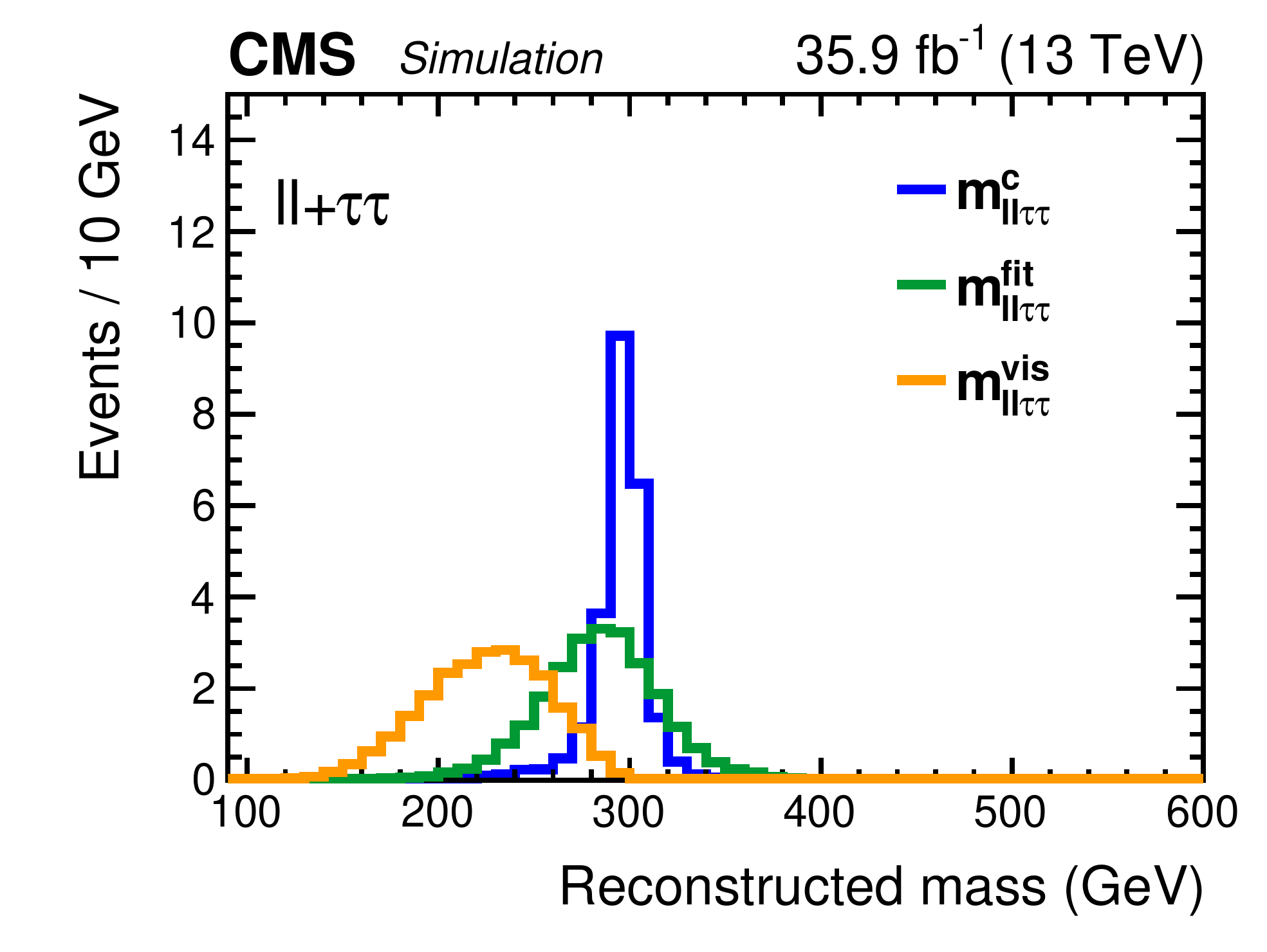}
 \caption{The distribution of the $\PA$ boson mass for the three studied mass reconstruction methods at 300\GeV: using only the visible decay products ($\mAvis$, orange), using the \textsc{svfit}
algorithm ($\mlltt^{\mathrm{fit}}$, green), and using the \textsc{svfit} algorithm with a mass constraint of 125\GeV for the Higgs boson ($\mlltt^{\mathrm{c}}$, blue).
The eight final states of the $\PA$ boson decay are combined for visualization purposes.}
 \label{fig:reconstructed_masses}
\end{figure}

The reconstructed mass of the $\PA$ boson candidate can be used to discriminate between signal- and background-like events.
Multiple reconstruction methods are considered and described below. The resulting mass distributions for the signal process ($\mA = 300\GeV$) are shown in Fig.~\ref{fig:reconstructed_masses}.
The shapes of the background distributions do not depend on the mass reconstruction method as strongly as the shape of the signal distribution.
The simplest reconstructed mass, $\mAvis$, uses only the visible decay products and combines the reconstructed
 $\PZ\to\ell\ell$ four-vector with the $\Ph\to\PGt\PGt$ four-vector based only on visible $\PGt$ decay products.
The resulting mass resolution for $\mAvis$ is approximately 15\% for an $\PA$ boson with a mass of 300\GeV in all final states.

The mass resolution of the reconstructed $\PA$ boson candidate can be significantly improved by
accounting for the neutrinos associated with the leptonic and hadronic tau decays.
We use the \textsc{svfit} algorithm~\cite{Bianchini:2016yrt} to estimate the mass of the Higgs boson, denoted as $\mtt^{\mathrm{fit}}$.
The \textsc{svfit} algorithm combines the \ptvecmiss with the four-vectors of both $\PGt$ candidates (electrons, muons,
or \tauh), resulting in an improved estimate of the four-vector of \Ph boson that is then used to obtain
a more accurate estimate of the $\PA$ boson candidate mass $\mlltt^{\mathrm{fit}}$.
The mass resolution of  $\mlltt^{\mathrm{fit}}$ is 10\% for an $\PA$ boson with a mass of 300\GeV.

To further improve the mass resolution, the measured mass of the Higgs boson (125\GeV) can be given as an input to the \textsc{svfit} algorithm.
This yields a constrained estimate of the four-vector of the \Ph boson,
which results in an even more precise estimate of the $\PA$ boson candidate mass, denoted as $\mlltt^{\mathrm{c}}$.
The resulting mass resolution of $\mlltt^{\mathrm{c}}$ is as good as 3\% at 300\GeV, which
improves the expected 95\% confidence level (\CL) model-independent limits by
approximately 40\% compared to using the visible mass of the $\PA$ boson $\mAvis$ as the discriminating variable.
Thus, we use $\mlltt^{\mathrm{c}}$ as the discriminating variable between the signal and the background processes for the final results.

\section{Event selection}\label{sec:selection}
Events are selected online using dilepton or single-lepton triggers targeting leptonic decays of the \PZ bosons.
The trigger and offline selection requirements for the \PZ boson decay modes are presented in Table~\ref{tab:inclusive_selection}.
Each lepton selected by the trigger is required to be geometrically matched to a corresponding lepton selected in the analysis.
The light leptons in an event are required to be separated from each
other by $\Delta R > 0.3$, while the \tauh candidates must be separated from each other and
from any other lepton by $\Delta R > 0.5$. The resulting event samples are made mutually
exclusive by discarding events that have additional identified
and isolated electrons or muons.
Small differences in trigger selection efficiencies are observed between data and simulation,
and are accounted for by applying corrections to the simulated events.

The nontriggering electrons and muons are required to have $\pt>10\GeV$, whereas \tauh candidates are required to have $\pt>20\GeV$.
The $\abs\eta$ constraints from detector geometry
are $\abs{\eta^{\Pe}}<2.5$, $\abs{\eta^{\Pgm}}<2.4$, and $\abs{\eta^{\tauh}}<2.3$ for
electrons, muons, and \tauh candidates, respectively. The $\abs\eta$ boundaries
are the same for both triggering and nontriggering electrons and muons.

\begin{table}[htbp]
\centering
\topcaption{Trigger and offline selection requirements for the different \PZ boson decay modes.
The events are selected using either dilepton triggers with lower-$\pt$ thresholds or single-lepton triggers with higher-$\pt$ thresholds.
The subscripts 1 and 2 indicate the higher- and lower-$\pt$ leptons associated with the \PZ boson, respectively.
\label{tab:inclusive_selection}
}
    \begin{tabular}{lll}
    \hline
     Decay channel           &       $\PZ\to\ell\ell$  trigger selection         & $\PZ\to\ell\ell$ offline selection   \\
    \hline
      $\PZ\to\Pe\Pe $     & $\left[\pt^{\Pe_{1}}>23\GeV~\&\,\pt^{\Pe_{2}}>12\GeV\right]$  &  $\left[\pt^{\Pe_{1}}>24\GeV~\&\,\pt^{\Pe_{2}}>13\GeV\right]$    \\
                   & or $\pt^{\Pe_{1}}>27\GeV$                   &  or $\left[\pt^{\Pe_{1}}>28\GeV~\&\,\pt^{\Pe_{2}}>10\GeV\right]$                   \\[\cmsTabSkip]
      $\PZ\to\Pgm\Pgm $    &  $\left[\pt^{\Pgm_{1}}>17\GeV~\&\,\pt^{\Pgm_{2}}>8\GeV\right]$ &  $\left[\pt^{\Pgm_{1}}>18\GeV~\&\,\pt^{\Pgm_{2}}>10\GeV\right]$   \\
                   &   or $\pt^{\Pgm_{1}}>24\GeV$                 & or $\left[\pt^{\Pgm_{1}}>25\GeV~\&\,\pt^{\Pgm_{2}}>10\GeV\right]$                     \\
    \hline
    \end{tabular}
\end{table}

The \PZ boson is reconstructed from a pair of opposite-charge, same-flavor light leptons that fulfills $60 < m_{\ell\ell} < 120\GeV$.
In case of multiple \PZ boson candidates, we choose the one with the mass closest to the \PZ boson mass.
Loose identification and isolation selection criteria are applied to the leptons associated to the \PZ boson
to maintain a high signal acceptance.
The leptons forming the \PZ boson candidate are required to pass the lepton identification, which has an efficiency of
90 ($>$99)\% for electrons (muons).
The muons must pass an isolation requirement of $I^{\Pgm}<0.25$, while a loose isolation requirement is already included
in the electron identification selection.

The leptons associated with the \Ph boson decay are required to have opposite charge.
In case of the $\Pe\tauh$, $\Pgm\tauh$, and $\Pe\Pgm$ decay channels, tighter selection criteria are applied to the light leptons
to decrease the background contributions from $\PZ+\text{jets}$ and other reducible
backgrounds. The specific signal selections detailed in Table~\ref{tab:signal_cuts},
including those chosen for the \tauh candidates,
were optimized to obtain the best signal sensitivity.
The isolation requirements are $I^{\Pe(\Pgm)}<0.15$ for electrons (muons) associated to a tau lepton decay.
Electrons from tau lepton decays need to pass the electron identification which has an efficiency of 80\%.
The \tauh candidates associated with the Higgs boson must satisfy the \tauh identification and isolation requirements which have an efficiency of 70\%.

The large \Ph boson mass leads to relatively high-$\pt$ decay
products compared to the lower \pt of jets misidentified as leptons from the $\PZ+\text{jets}$ background process.
This background process is suppressed by selecting events based on the scalar \pt sum of the visible decay
products of the Higgs boson, $\LTH$.
In the $\ell\ell+\tauh\tauh$ final states, which have a
larger relative ratio of reducible to irreducible backgrounds, events with $\LTH > 60\GeV$ are selected.

The signal events contain no $\cPqb$ jets ($\Pg\Pg\to\PA$), or only $\cPqb$ jets with a relatively soft \pt distribution ($\bbbar\PA$).
We suppress the contributions from background processes, especially $\ttbar$ and $\ttbar\PZ$, by discarding all events with one or more $\cPqb$-tagged jets with $\pt>20\GeV$ (``$\cPqb$ jet veto'') without significantly
reducing the signal selection efficiency. The total acceptance for the $\Pg\Pg\to\PA$ ($\bbbar\PA$) signal events with $\mA = 300\GeV$ is 3.9 (3.0)\%.
The fraction of $\Pg\Pg\to\PA$ signal events lost due to the $\cPqb$ jet veto is negligible, while for the $\bbbar\PA$ process approximately 17\% of events are removed with this selection.

The sensitivity of the analysis is improved by reducing the number of background events using additional information regarding the Higgs boson candidate.
The constrained Higgs boson candidate four-vector, as estimated with the \textsc{svfit} algorithm, is used to reconstruct the $\PA$ boson mass, as described in Section~\ref{sec:reconstruction}.
By removing the mass constraint from the \textsc{svfit} algorithm, the most likely mass of the Higgs boson candidate $\mtt^{\mathrm{fit}}$
provides significant discrimination between reducible backgrounds,
which have a broad distribution due to their nonresonant nature, and the signal processes, which have a resonance present at 125\GeV.
Moreover, the dominant irreducible background from $\PZ\PZ\to4\ell$ ($\cPq\cPq\to\PZ\PZ$ and $\Pg\Pg\to\PZ\PZ$) is suppressed, because
for this background the $\mtt^{\mathrm{fit}}$ distribution is concentrated near the \PZ boson mass in contrast to the signal.
The signal sensitivity is increased by an additional $20\%$ by requiring $\mtt^{\mathrm{fit}}$ to be within 90--180\GeV.

\begin{table}[htbp]
\centering
\topcaption{Kinematic selection requirements for each $\PA$ boson decay channel,
applied on top of the looser selections and $\cPqb$ jet veto described in the text.
The efficiency of the identification (and isolation) requirement for a given lepton type is labeled $\epsilon_{\mathrm{id.}}^{\ell}$.
The leptons assigned to the Higgs boson are required to have opposite charge.
To increase the sensitivity, we require $\mtt^{\mathrm{fit}}$ to be within 90--180\GeV.
In the $\ell\ell+\tauh\tauh$ channel, we additionally require $\LTH>60\GeV$, where $\LTH$ is the scalar \pt sum of the visible decay products of the Higgs boson.
\label{tab:signal_cuts}
}
\cmsTable{
\begin{tabular}{ccc}
    \hline
  Channel         & \PZ boson selection    &  \Ph boson selection     \\
\hline
  $\ell\ell+\Pe\tauh$   &    \multirow{4}{*}[-0.0cm]{\hspace{-0.01cm} $\left. \vphantom{\begin{tabular}{c}4\\4\\4\\4\end{tabular}}\right\{$ \hspace{7cm} $\left. \vphantom{\begin{tabular}{c}4\\4\\4\\4\end{tabular}}\right\}$ }      &  {\footnotesize$\epsilon_{\mathrm{id.}}^{\Pe}=80\%$, $I^{\Pe}<0.15$}, {\footnotesize$\epsilon_{\mathrm{id.+iso.}}^{\tauh}=70\%$}  \\
  $\ell\ell+\Pgm\tauh$  & Opposite-charge, same-flavor light leptons    &  {\footnotesize $\epsilon_{\mathrm{id.}}^{\Pgm}>99\%$, $I^{\Pgm}<0.15$ }, {\footnotesize$\epsilon_{\mathrm{id.+iso.}}^{\tauh}=70\%$}                                       \\
  $\ell\ell+\tauh\tauh$ &     $60 < m_{\ell\ell} < 120\GeV$    &  {\footnotesize$\epsilon_{\mathrm{id.+iso.}}^{\tauh}=70\%$, $\LTH>60\GeV$}   \\
  $\ell\ell+\Pe\Pgm$ &     &         {\footnotesize$\epsilon_{\mathrm{id.}}^{\Pe}=80\%$, $I^{\Pe}<0.15$},  {\footnotesize$\epsilon_{\mathrm{id.}}^{\Pgm}>99\%$, $I^{\Pgm}<0.15$}   \\
    \hline
\end{tabular}
}
\end{table}

\section{Background estimation}\label{sec:background_estimation}
The irreducible backgrounds ($\PZ\PZ\to4\ell$, $\ttbar\PZ$, $\PW\PW\PZ$, $\PW\PZ\PZ$, $\PZ\PZ\PZ$) and the production of the 125\GeV Higgs boson via the processes predicted by the SM are estimated from simulation.
They are scaled by their theoretical cross sections calculated at the highest order available, and the processes producing the 125\GeV Higgs boson are also scaled by their most accurate branching fractions~\cite{deFlorian:2016spz}.

The reducible backgrounds, which have at least one jet misidentified as an electron,
muon, or \tauh candidate, are estimated from data. In this analysis the dominant reducible contributions come
from the $\ttbar$, $\PZ+\text{jets}$, and $\PW\PZ+\text{jets}$ processes which produce jets misidentified as $\Pgt$ candidates.
The estimation of the reducible background contribution is performed with a so-called ``fake rate method'' which is
based on measuring the misidentification rates, \ie, probabilities to misidentify a jet as a lepton.
Events with $\Pgt$ candidates failing the signal region identification and isolation criteria are used along
with the misidentification rates to estimate the
contribution from the reducible background in the signal region.

In total three different event samples are used to estimate the contribution from the reducible background processes.
First, the misidentification rates are estimated in event samples independent from the signal region. This region is called a ``measurement region''.
To understand to which extent the measured misidentification rates describe the jets misidentified as leptons in the signal region,
closure tests comparing the observed and the estimated reducible background yields are performed in yet another region (``validation region'').
The validation region is required to be independent from the signal and the measurement regions.
The closure tests are used to derive systematic uncertainties to account for possible differences between the true and the estimated reducible background yields in the signal region.
Finally, the misidentification rates are applied in an ``application region'', formed by events that fail the identification and isolation criteria required in the signal region.

In this analysis we use a sample of $\PZ+\text{jet}$ events to estimate the misidentification rates.
The estimation of misidentification rates relies on reconstructing an opposite-charge, same-flavor lepton
pair compatible with a \PZ boson, and requiring one additional loosely defined lepton (electron, muon, or \tauh candidate).
The requirements on the leptons associated with the \PZ boson are the same as defined in Section~\ref{sec:selection}, but they must
fulfill a more stringent dilepton mass requirement, $81.2 < m_{\ell\ell} < 120\GeV$.
After reconstructing the $\PZ\to\ell\ell$ decay, the jet-to-lepton
misidentification rate is estimated by applying the
lepton identification algorithm to the additional loosely defined lepton in the event.
The misidentification rates are measured in different bins of lepton \pt, and are further split between
reconstructed decay modes for the \tauh candidate, and for muons and electrons in bins of lepton $\eta$, based on the barrel and endcap regions.
The events where the $\Pgt$ candidates arise from genuine tau leptons, electrons, or muons and not jets,
primarily from the $\PW\PZ$ process, are estimated from simulation and subtracted
from data so that the misidentification rates are measured for genuine hadronic jets only.
The obtained misidentification rates for electrons (muons) are $<$5 (10)\% in barrel and endcap regions for lepton $\pt>10\GeV$,
whereas for \tauh candidates the misidentification rates vary between $15$--$30\%$ for \tauh candidate $\pt>20\GeV$ depending on the decay mode.

The measured misidentification rates are validated in another region that consists of events with a \PZ boson
candidate and two additional loosely defined leptons. To ensure that the validation region is not contaminated with signal events or
irreducible background contributions, the two additional leptons are required to have the same charge.
Modest differences in observed versus predicted reducible background yields are observed.
These differences are accounted for by assigning a systematic uncertainty in the yield, taken to be 40\% which is conservative enough to cover the observed nonclosure.
This uncertainty is uncorrelated between the Higgs boson decay channels
resulting in four uncertainties tied to $\ell\ell+\Pe\tauh$, $\ell\ell+\Pgm\tauh$, $\ell\ell+\tauh\tauh$, and $\ell\ell+\Pe\Pgm$ channels.
Further studies confirmed that the final results of this analysis are not sensitive to the exact magnitude of this systematic uncertainty.

To estimate the reducible background contribution in the signal region,
we apply a weight on data events where
either one or both of the $\Pgt$ candidates associated
to the Higgs boson fail the identification and isolation criteria. These data events form the application region.

Events with exactly one object failing the identification and isolation criteria receive a weight $f/(1-f)$,
where $f$ is the misidentification rate for the particular type of lepton. As such, this weight
includes the contribution from the $\PW\PZ+\text{jets}$ process,
where we expect one genuine lepton and one jet misidentified as a lepton in addition to the \PZ boson candidate.
Also $\ttbar$ and $\PZ+\text{jets}$ processes are accounted for by the weight as either of the two jets
can pass the identification and isolation criteria even if neither of them is a genuine lepton.
As a result, the weight introduces double counting of events from $\ttbar$ and $\PZ+\text{jets}$ processes.

To remove the double-counted events from $\ttbar$ and $\PZ+\text{jets}$ processes, we define a weight with a negative sign that is given
for events with both objects failing the identification and isolation criteria, namely $-f_1f_2/[(1-f_1)(1-f_2)]$.
This subtraction, however, introduces increased statistical uncertainties on the estimated yield of the reducible background.

The statistical uncertainties can be controlled by taking the shape of the $\mlltt^{\mathrm{c}}$ distribution of the reducible background contribution from data
in another region with negligible signal and irreducible background contributions. This region is defined similarly to the signal region
but with same-sign $\Pgt$ candidates passing relaxed identification and isolation criteria, yielding a higher number of events available for the shape estimation.
This results in a smoother shape of the $\mlltt^{\mathrm{c}}$ distribution, which is normalized to the estimated yield of the reducible background contribution in the signal region.

An alternative approach to estimate the reducible background contribution was studied to reduce the statistical uncertainties and to cross check the
results obtained using the nominal method.
Instead of using the same-sign data events for the shape of the $\mlltt^{\mathrm{c}}$ distribution, the statistical uncertainties can be
reduced considerably by giving a suitable nonzero weight only for events with both candidates failing the selection criteria, \ie, by
estimating only the contribution from the $\ttbar$ and $\PZ+\text{jets}$ processes by using the misidentification rate method.
The contribution from events with a single object failing the identification and isolation criteria is predicted from simulation,
removing the double counting present in the nominal method.
As a result, this alternative approach requires using a weight with a positive sign ($f_1f_2/[(1-f_1)(1-f_2)]$).
Since the statistical uncertainties are smaller, the shape of the $\mlltt^{\mathrm{c}}$ distribution is taken from the same events that provide the estimated yield of the reducible background.
The results of the cross-check show that the two methods yield consistent expected 95\% \CL model-independent limits.

To cross check the measured misidentification rates, we performed an additional measurement using a sample of $\PZ+2\ \text{jets}$ events.
In this cross-check, the measurement region partially overlaps with the aforementioned validation region, as in both cases the two
lepton candidates are required to have the same charge.
The amount of overlap between the measurement and validation regions depends on the lepton type and the decay channel of the Higgs boson.
The rates are measured in bins of lepton \pt, and are separated by the reconstructed decay mode of the \tauh candidates.
Unlike above, the misidentification rates are not split in bins of lepton $\eta$ for electrons and muons.
The measured misidentification rates result in a reducible background yield and shape that are
compatible with the reducible background estimation obtained with the nominal misidentification rate measurement used in this analysis.

\section{Systematic uncertainties}\label{sec:systematics}
All systematic uncertainties considered in the analysis are summarized in Table~\ref{tab:uncertainties}.
Different uncertainties are treated as uncorrelated, and each uncertainty is assumed correlated between different processes and final states, unless otherwise mentioned below.

The overall uncertainty in the \tauh identification and isolation efficiency for genuine \tauh
leptons is 5\%~\cite{Sirunyan:2018pgf}, which has been measured with a tag-and-probe method in $\PZ\to\PGt\PGt$ events.
An uncertainty of 1.2\% in the visible energy scale of genuine \tauh candidates affects both the distributions and yields of the signals and backgrounds. It is
uncorrelated across the 1-prong, 1-prong+$\PGpz\mathrm{s}$, and 3-prong decay modes.

The uncertainties in the electron and muon identification and isolation efficiencies lead to a normalization uncertainty of 2\% for either electrons or muons.
The uncertainty in the trigger efficiency results in a normalization uncertainty of 2\% for both electron and muon triggers.
In all channels, the effect of the uncertainty in the electron and muon energy scales is negligible.

The normalization uncertainty related to vetoing events with a $\cPqb$-tagged jet is 4.5\%
for the background processes with heavy-flavor jets (from charm or bottom quarks), \ie, $\ttbar$, $\ttbar\PZ$, and $\ttbar\PW$.
All other processes, including the signal process, are dominated by light-flavor or gluon jets and their normalization uncertainty is 0.15\%.

The normalization uncertainties related to the choice of PDFs, and the renormalization and factorization (RF) scales,
affecting the acceptance of the dominant background processes, are estimated from simulation separately for each process.
The uncertainty from the RF scales is determined by varying one scale at a time by factors of
0.5 and 2.0, and calculating the change in process acceptance.
Combining the RF scale uncertainties with the PDF set uncertainty~\cite{Butterworth:2015oua} for the
$\cPq\cPq\to\PZ\PZ$ process leads to an uncertainty of 4.8\%.
The inclusive uncertainty for $\PZ\Ph$ production related to the PDFs amounts to 1.6\%,
whereas the uncertainty for the variation of the RF scales is 3.8\%~\cite{deFlorian:2016spz}. For the subleading \Ph boson processes $\PW\Ph$,
$\Pg\Pg\to\Ph\to\PZ\PZ$, and $\ttbar\Ph$ the inclusive uncertainties related to the PDFs amount to
1.9, 3.2, and 3.6\% and the uncertainties for the variation of the RF scales are 0.7, 3.9, and 7.5\%, respectively~\cite{deFlorian:2016spz}.

For the $\Pg\Pg\to\PZ\PZ$ process, there is a 10\% uncertainty in the NNLO cross section estimate used in the
analysis, which covers the PDF, RF scale uncertainties, and the uncertainty on the strong coupling constant.
An additional 10\% uncertainty is included to account for the assumptions used to estimate the NNLO cross section~\cite{Sirunyan:2019twz}.
The uncertainties in the cross section of the rare $\ttbar\PZ$, $\ttbar\PW$, and triboson processes amount to 25\%~\cite{Sirunyan:2017uzs}.

The last theoretical uncertainty applied in this analysis is the uncertainty in the theoretical calculations
of the SM $\Ph\to\Pgt\Pgt$ branching fraction. This uncertainty of 2\%~\cite{deFlorian:2016spz} is applied to both the $\Pg\Pg\to\PA$ and $\bbbar\PA$ signal samples as well as all backgrounds that include the $\Ph\to\Pgt\Pgt$ process.

Normalization uncertainties in the misidentification rates arising from the subtraction of
prompt lepton contribution estimated from simulation are taken into account and propagated to
the yield of the reducible background mass distributions.
The shape of the $\mlltt^{\mathrm{c}}$ distribution of the reducible background is estimated from data in
a region where the $\Pgt$ candidates have the same charge and pass relaxed isolation conditions.
Therefore, the statistical uncertainties in the misidentification rates do not have an impact on the shape of the $\mlltt^{\mathrm{c}}$ distribution.
As discussed in Section~\ref{sec:background_estimation}, an additional uncertainty is applied based on
the results of the closure tests comparing the differences between the observed and the estimated reducible background yields.
The uncertainty in the yield is taken to be 40\%, and is considered uncorrelated
across the $\ell\ell+\Pe\tauh$, $\ell\ell+\Pgm\tauh$, $\ell\ell+\tauh\tauh$, and $\ell\ell+\Pe\Pgm$ channels.

The $\ptvecmiss$ scale uncertainties~\cite{Sirunyan:2019kia}, which are computed
event-by-event, affect the normalization of various processes
as well as their distributions through the propagation of these
uncertainties to the di-tau masses $\mtt^{\mathrm{fit}}$ and $\mtt^{\mathrm{c}}$. The
$\ptvecmiss$ scale uncertainties arising from unclustered energy deposits in the
detector come from four independent sources related to the tracker, ECAL, hadron calorimeter,
and forward calorimeters. The $\ptvecmiss$ scale
uncertainties related to the uncertainties in the jet energy scale measurement,
which affect the $\ptvecmiss$ calculation, are taken into
account as a separate uncertainty.

Uncertainties related to the finite number of simulated events are taken into account using the Barlow-Beeston-lite method~\cite{Barlow:1993dm}. They
are considered for all bins of the background distributions used to extract the results.
They are uncorrelated across different samples, and across bins of a single distribution.
Finally, the uncertainty in the integrated luminosity amounts to 2.5\%~\cite{CMS-PAS-LUM-17-001}.

\begin{table}[!ht]
\centering
\topcaption{Sources of systematic uncertainty. The sign $\dagger$ marks the uncertainties that
affect both the shape and normalization of the final $\mlltt^{\mathrm{c}}$ distributions. Uncertainties that only affect the normalizations have no marker.
For the shape and normalization uncertainties, the magnitude column lists an approximation of the associated change in the normalization of the affected processes.
}
\newcolumntype{x}{D{,}\text{{--}}{2.2}}
\cmsTable{
\begin{tabular}{lll}
    \hline
Source of uncertainty & Process & Magnitude\\
\hline
 \tauh id.\ \& isolation          & All simulated processes                  & 5\% \\
 \tauh energy scale$^{\dagger}$ (1.2\% energy shift)     & All simulated processes              & $<$2\% \\
 $\Pe$ id.\ \& isolation      & All simulated processes               & 2\% \\
 $\Pe$ trigger               & All simulated processes      & 2\% \\
 $\Pgm$ id.\ \& isolation    & All simulated processes              & 2\%  \\
 $\Pgm$ trigger            & All simulated processes         & 2\% \\
 $\cPqb$ jet veto                & All simulated processes                             & 4.5\% heavy flavor, 0.15\% light flavor or gluon \\
 $\cPq\cPq\to\PZ\PZ$ theoretical uncertainty          & $\cPq\cPq\to\PZ\PZ$          & 4.8\% \\
 PDF set uncertainty &     $\PZ\Ph$, $\PW\Ph$, $\Pg\Pg\to\Ph\to\PZ\PZ$, and $\ttbar\Ph$      & Varies from 1.6 to 3.6\% (see text)\\
 RF scale uncertainty &    $\PZ\Ph$, $\PW\Ph$, $\Pg\Pg\to\Ph\to\PZ\PZ$, and $\ttbar\Ph$       & Varies from 0.7 to 7.5\% (see text)\\
 $\Pg\Pg\to\PZ\PZ$ theoretical uncertainty     & $\Pg\Pg\to\PZ\PZ$            & 10\% \\
 $\Pg\Pg\to\PZ\PZ$ NNLO cross section estimation assumptions & $\Pg\Pg\to\PZ\PZ$  & 10\% \\
 $\ttbar\PZ$  theoretical uncertainty  & $\ttbar\PZ$       & 25\% \\
 $\ttbar\PW$  theoretical uncertainty  & $\ttbar\PW$       & 25\% \\
 Triboson  theoretical uncertainty    & Triboson     & 25\% \\
 Theoretical uncertainty on $\mathcal{B}(\Ph\to\PGt\PGt)$ & Signal, $\PZ\Ph$, and $\PW\Ph$  & $<$2\% \\
 Reducible background uncertainties:     & Reducible background           &   \\
 $\,\,\,\,\,\,\,\,$$\Pe$ prompt lepton subtraction  &         & $<$12\% in $\ell\ell+\Pe\Pgm$, $<$1\% in $\ell\ell+\Pe\tauh$  \\
 $\,\,\,\,\,\,\,\,$$\Pgm$ prompt lepton subtraction  &         & $<$16\% in $\ell\ell+\Pe\Pgm$, $<$1.5\% in $\ell\ell+\Pgm\tauh$  \\
 $\,\,\,\,\,\,\,\,$$\tau$ prompt lepton subtraction  &         & $<$3.5\% in $\ell\ell+\Pe\tauh$ and $\ell\ell+\Pgm\tauh$, $<$1\% in $\ell\ell+\tauh\tauh$  \\
 $\,\,\,\,\,\,\,\,$Normalization         &            & 40\% in $\ell\ell+\Pe\tauh$, $\ell\ell+\Pgm\tauh$, $\ell\ell+\tauh\tauh$, and $\ell\ell+\Pe\Pgm$  \\
 $\ptvecmiss$ energy scale$^{\dagger}$  & All simulated processes            & $<$2\%  \\
 Limited number of events      & All background processes                      & Statistical uncertainty in individual bins \\
 Integrated luminosity     & All simulated processes                         & 2.5\% \\
    \hline
\end{tabular}
\label{tab:uncertainties}
}
\end{table}

\section{Results}\label{sec:results}
We use the reconstructed pseudoscalar Higgs boson mass, $\mlltt^{\mathrm{c}}$, as the discriminating variable between the signal and the background processes.
The results are based on a simultaneous binned likelihood fit of the reconstructed mass distributions in the eight final states.
The eight final states are each fit as separate distributions in the simultaneous
fit. They are combined together for visualization purposes only. Nuisance parameters, representing the systematic uncertainties, are profiled in the fit.
Even though the studied signal mass range is 220--400\GeV, the distribution of the reconstructed mass $\mlltt^{\mathrm{c}}$ covers the mass range 200--600\GeV,
as the additional information on the background distributions is used to constrain the corresponding parameters in the simultaneous fit.
 When displaying the results, background processes are grouped as follows:
``$\Ph(125\GeV)$'' includes all processes with the SM Higgs boson (including $\Pg\Pg\to\Ph\to\PZ\PZ\to4\ell$);
``$\PZ\PZ\to4\ell$'' includes events from $\cPq\cPq\to\PZ\PZ$ and $\Pg\Pg\to\PZ\PZ$ processes;
``Other'' includes events from triboson, $\ttbar\PZ$, and $\ttbar\PW$ production; and ``Reducible'' includes the reducible background contribution.

\begin{figure}[h!]
\centering
  \includegraphics[width=0.45\textwidth]{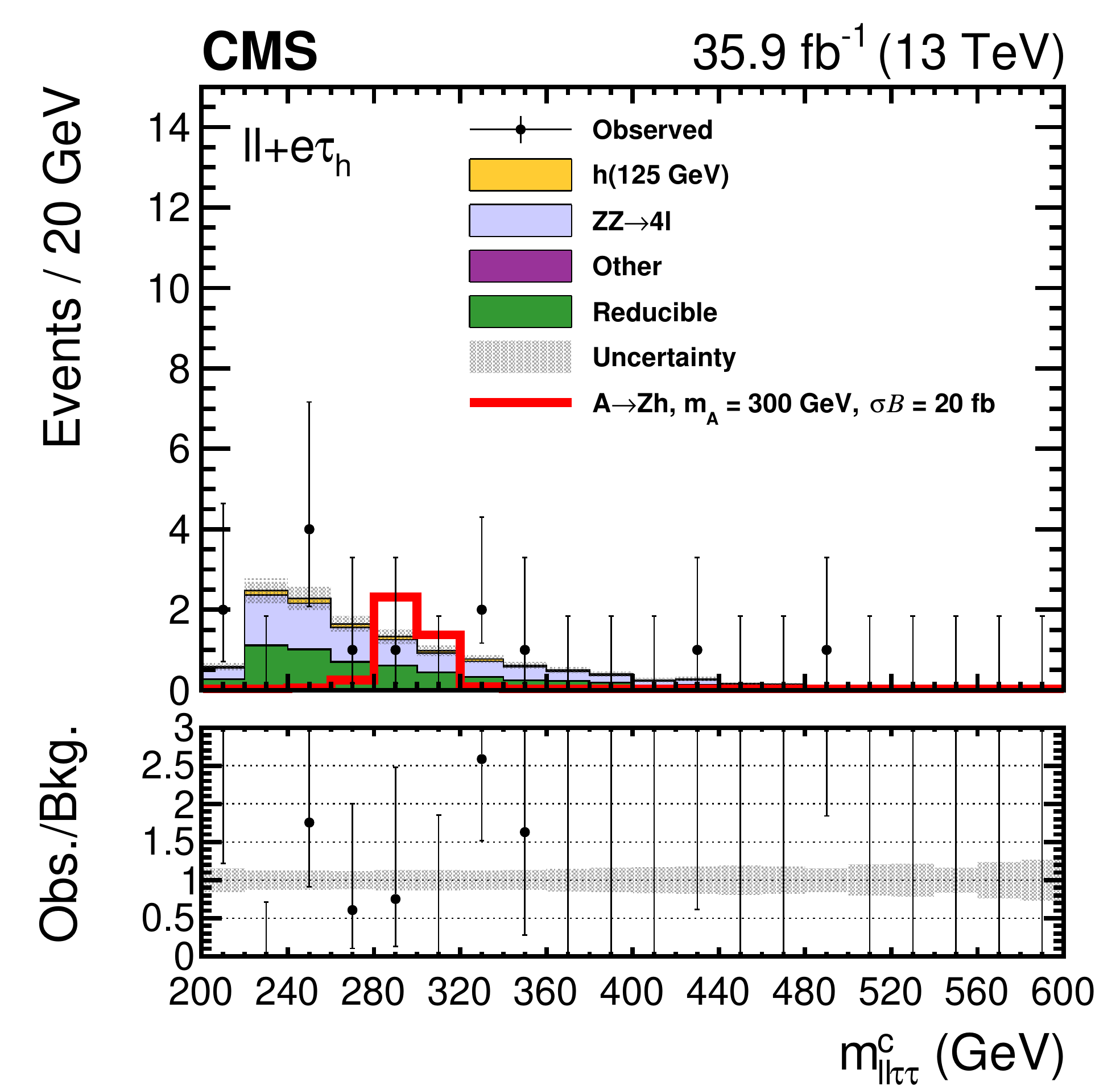}
  \includegraphics[width=0.45\textwidth]{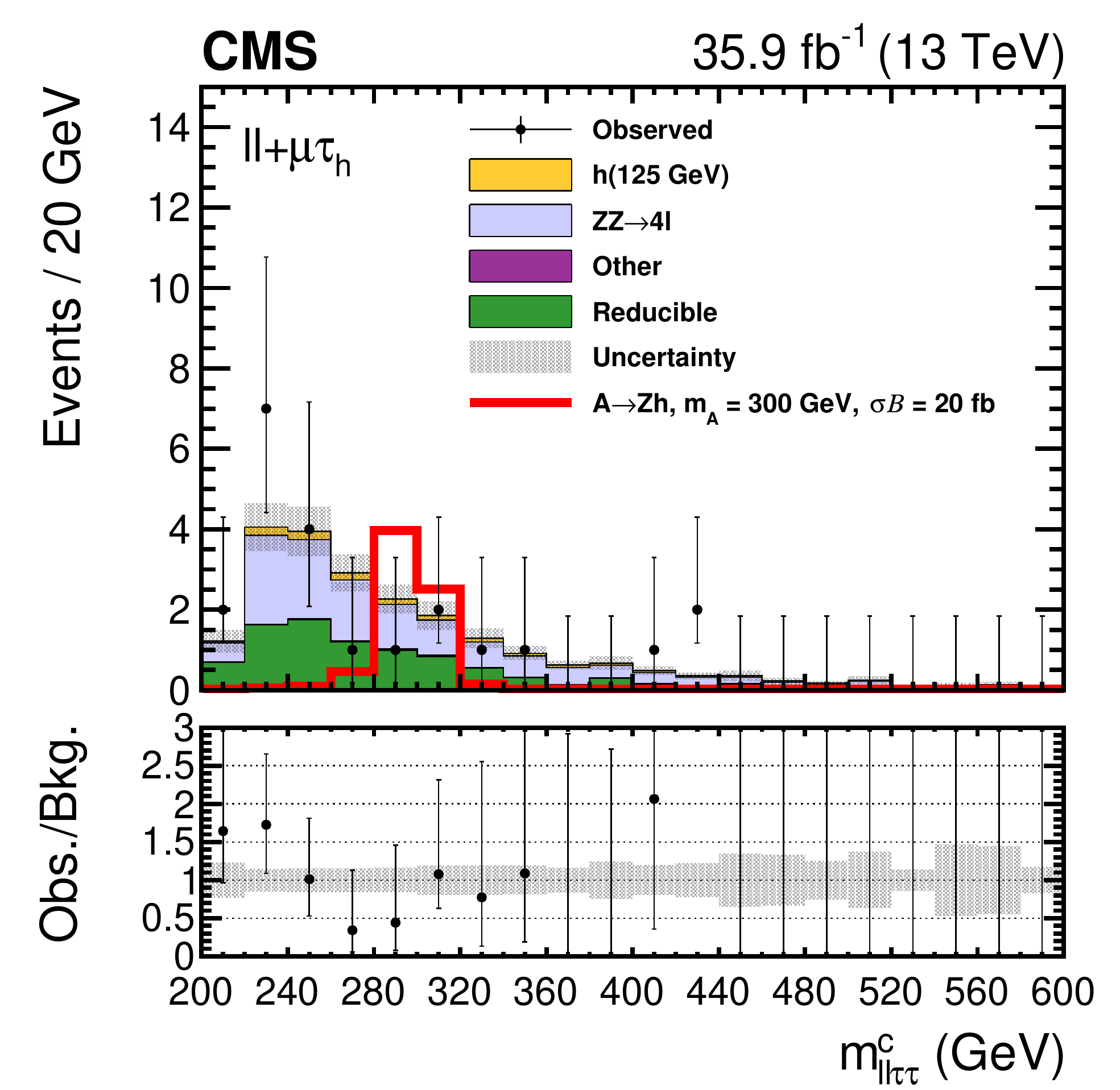}
  \includegraphics[width=0.45\textwidth]{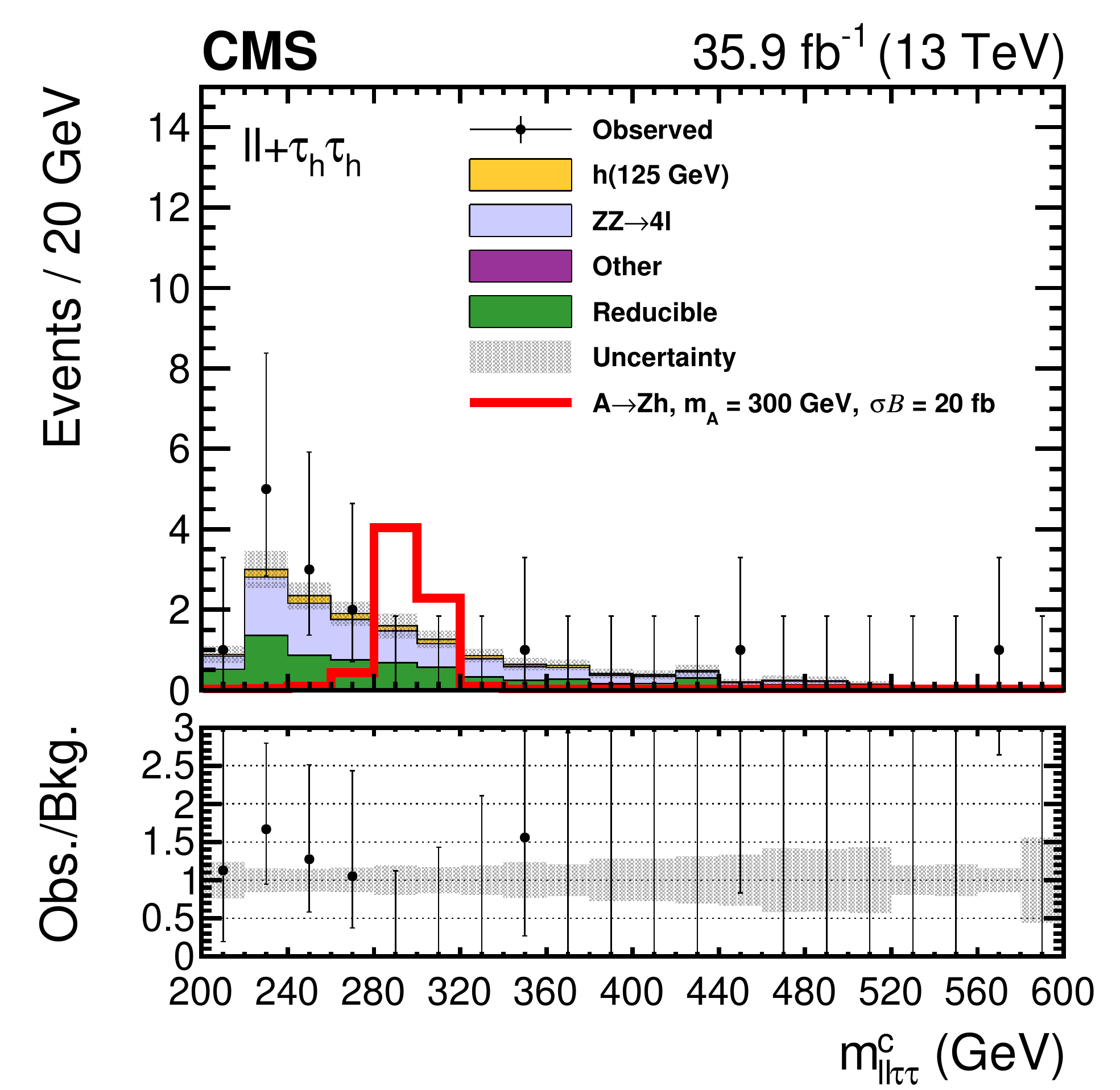}
  \includegraphics[width=0.45\textwidth]{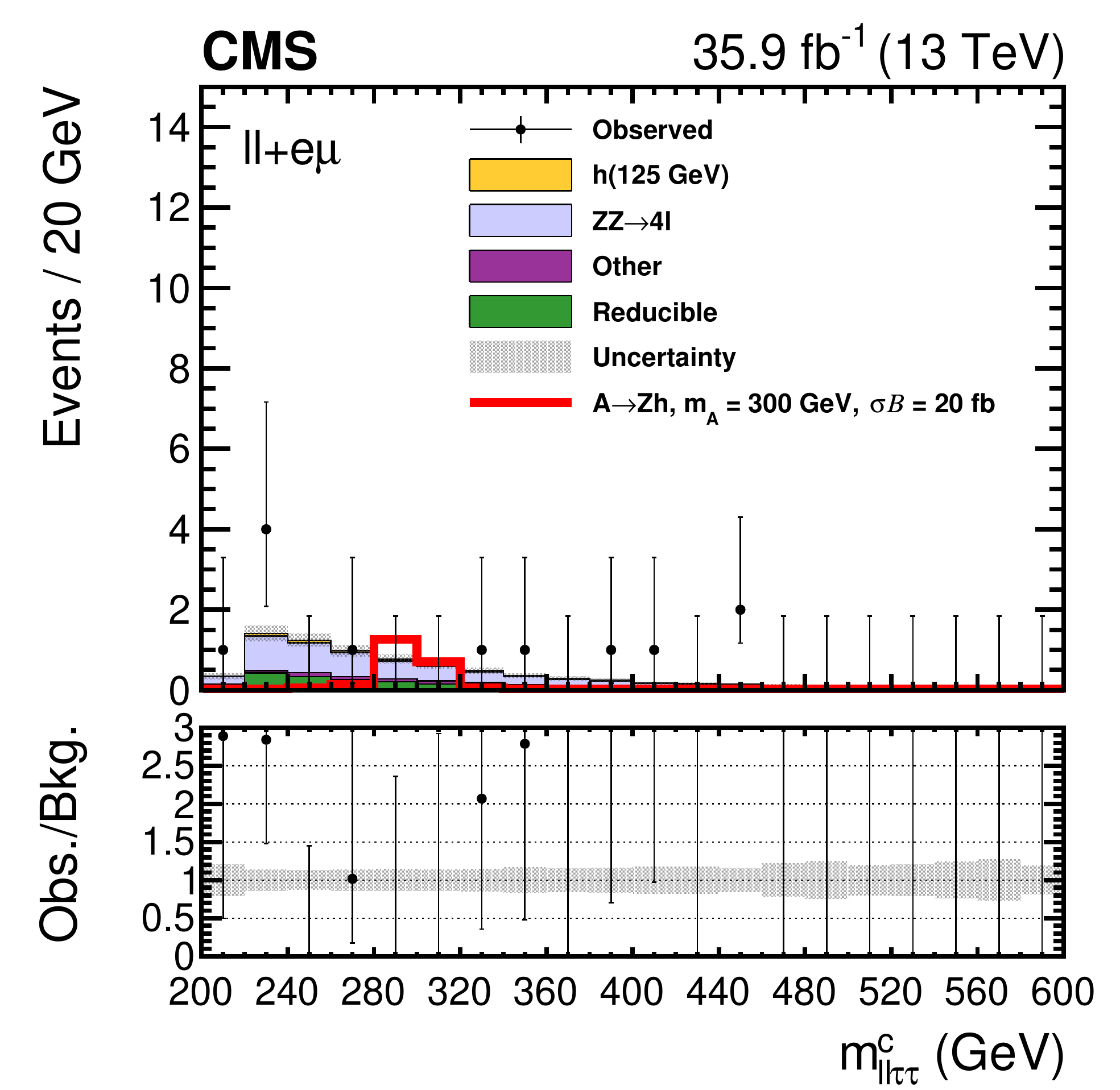}
 \caption{The reconstructed mass $\mlltt^{\mathrm{c}}$ distributions and uncertainties after a background-only fit
  for the $\ell\ell+\Pe\tauh$ (upper left), $\ell\ell+\Pgm\tauh$ (upper right),
  $\ell\ell+\tauh\tauh$ (lower left), and $\ell\ell+\Pe\Pgm$ (lower right) channels.
  In all cases the two decay channels of the \PZ boson
   are included as separate distributions in the simultaneous fit; combining them together is for visualization purposes only.
  The uncertainties include both statistical and systematic components.
  The expected contribution from the $\PA\to\PZ\Ph$ signal process is shown for a pseudoscalar Higgs boson
  with $\mA = 300\GeV$ with the product of the cross section and branching fraction of 20\unit{fb} and is for illustration only.
 }
 \label{fig:azh_results_LLXX}
\end{figure}

\begin{figure}[h!]
\centering
  \includegraphics[width=0.65\textwidth]{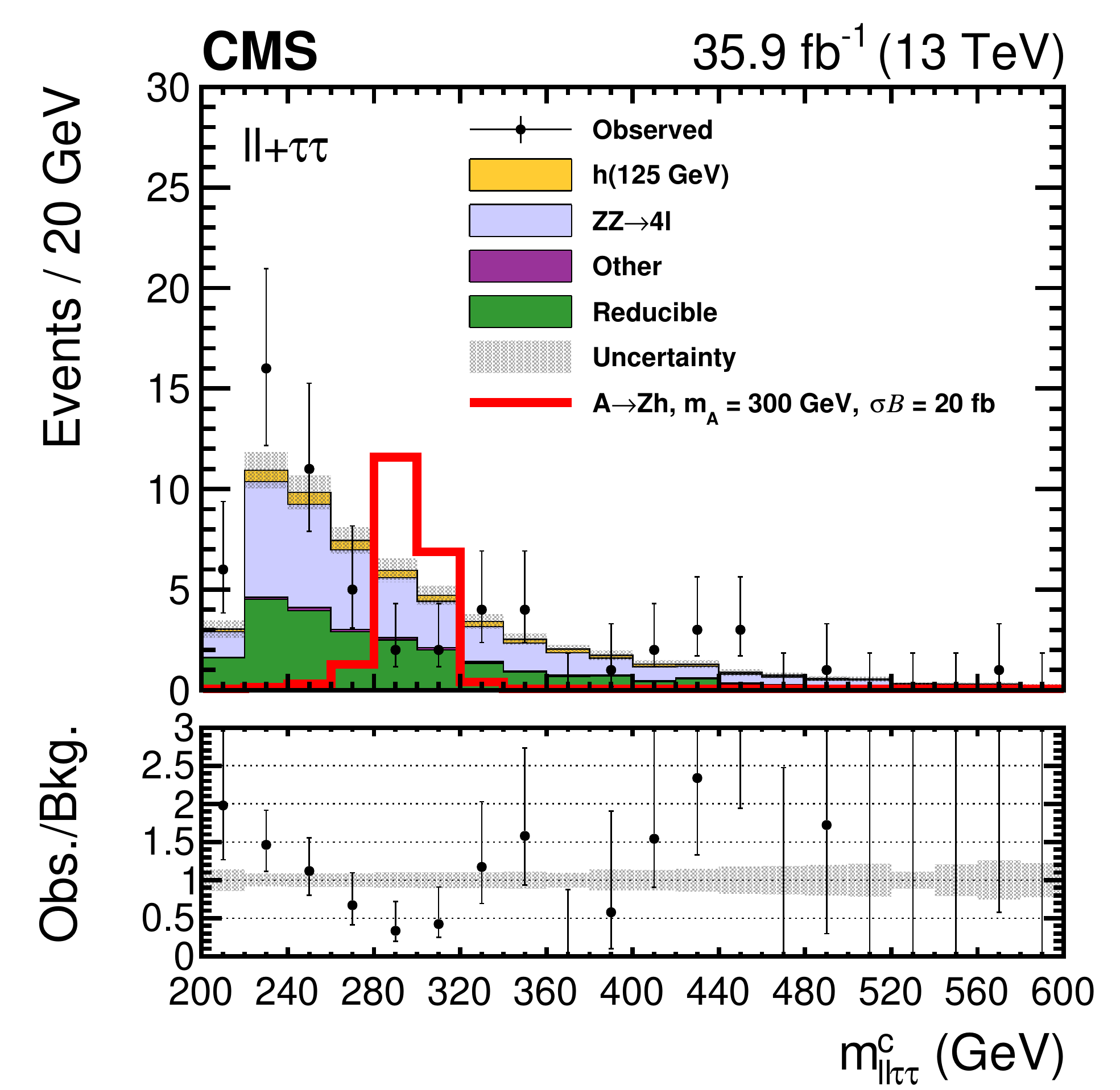}
 \caption{The reconstructed mass $\mlltt^{\mathrm{c}}$ distribution and uncertainties after a background-only fit in all eight final states.
  The final states are included as separate distributions in the simultaneous fit; combining them together is for visualization purposes only. The uncertainties include both statistical and systematic components.
  The expected contribution from the $\PA\to\PZ\Ph$ signal process is shown for a pseudoscalar Higgs boson
  with $\mA = 300\GeV$ with the product of the cross section and branching fraction of 20\unit{fb}
  and is for illustration only.
 }
 \label{fig:azh_results_All}
\end{figure}

The $\mlltt^{\mathrm{c}}$ distributions are shown in Fig.~\ref{fig:azh_results_LLXX} for each of the four \Ph boson decay channels, adding the $\PZ\to\ell\ell$ channels together,
and in Fig.~\ref{fig:azh_results_All} for all eight final states together. The distributions are shown after a background-only fit to data and include both
statistical and systematic uncertainties.
No excess above the standard model background expectations is observed in data.
The predicted signal and background yields, as well as the number of observed events, are given in Table~\ref{tab:sb_zh}
for each of the four $\PZ\Ph$ channels.

\begin{table}
\centering
\topcaption{Background and signal expectations together with the numbers of observed
events, for the signal region distributions after a background-only fit.
The expected contribution from the $\PA\to\PZ\Ph$ signal process is given for a pseudoscalar Higgs boson
  with $\mA = 300\GeV$ with the product of the cross section and branching fraction of 20\unit{fb}.
The background uncertainty accounts for all sources of background uncertainty,
systematic as well as statistical, after the simultaneous fit.
}
\label{tab:sb_zh}
\cmsTable{
\newcolumntype{x}{D{,}{\,\pm\,}{5.5}}
\begin{tabular}{lxxxx}
    \hline
Process & \multicolumn{1}{c}{$\ell\ell+\Pe\tauh$} &  \multicolumn{1}{c}{$\ell\ell+\Pgm\tauh$} &  \multicolumn{1}{c}{$\ell\ell+\tauh\tauh$} &  \multicolumn{1}{c}{$\ell\ell+\Pe\Pgm$} \\
\hline
\Ph(125\GeV)                                  & 0.77, 0.02    &   1.39, 0.03   &  1.28, 0.04   &   0.45, 0.01    \\
$\PZ\PZ\to4\ell$                                  & 6.48, 0.13     &   11.38, 0.25    &  7.59, 0.20    &   4.57, 0.09     \\
Other                                     & 0.10, 0.01     &   0.24, 0.02  &  0.04, 0.01  &   0.69, 0.04      \\
Reducible                                 & 5.52, 0.42     &   9.12, 0.93    &  6.68, 0.65    &   2.04, 0.24     \\
Total background                         & 12.88, 0.45     &  22.13, 0.94      &  15.58, 0.68     &   7.74, 0.28       \\[\cmsTabSkip]
$\PA\to\PZ\Ph$, $\mA=300\GeV$, $\sigma \mathcal{B}=20\unit{fb}$ & 4.13, 0.18   &   7.32, 0.30    &  7.01, 0.40    &   2.26, 0.10     \\[\cmsTabSkip]
Observed &  \multicolumn{1}{c}{13} &  \multicolumn{1}{c}{22} &  \multicolumn{1}{c}{14} &  \multicolumn{1}{c}{12}  \\
    \hline
\end{tabular}
}
\end{table}

\begin{figure}[h!]
\centering
  \includegraphics[width=0.65\textwidth]{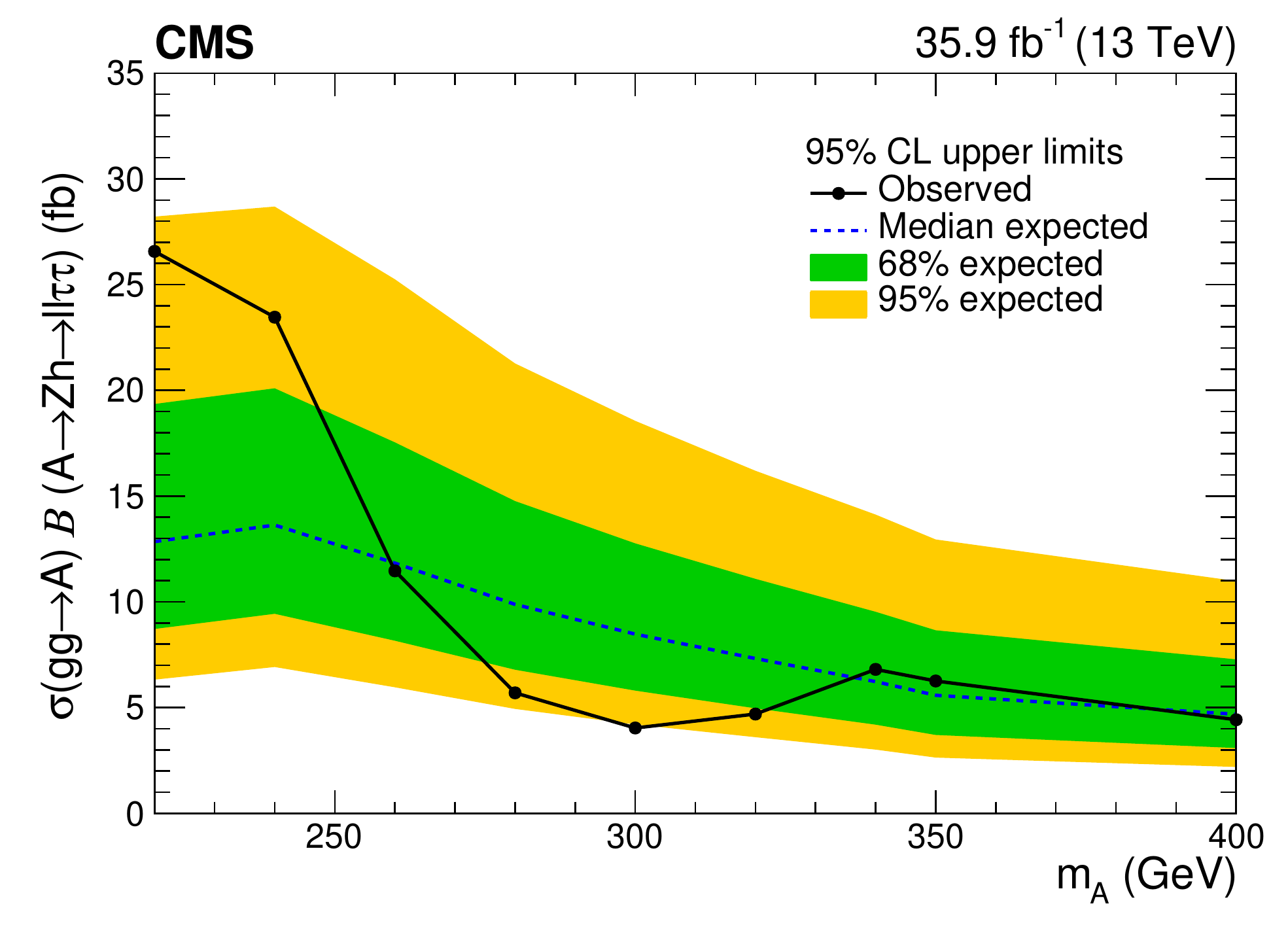}
 \caption{
  The expected and observed 95\% \CL model-independent upper limits on the product of the cross section and branching fraction
$\sigma(\Pg\Pg\to\PA)\mathcal{B}(\PA\to\PZ\Ph\to\ell\ell\PGt\PGt)$ are shown. The green (yellow) band corresponds to the 68 (95)\% confidence
intervals for the expected limit.
 }
 \label{fig:limits_model_indep}
\end{figure}

Upper limits at 95\% \CL~\cite{CLS2, CLS1} are set in multiple scenarios.
An asymptotic approximation of the modified frequentist \CLs method~\cite{CLS2, CLS1, CMS-NOTE-2011-005, Cowan:2010js} is used when calculating the 95\% \CL upper limits.
Model-independent limits are set on the product of the cross section and branching fraction,
$\sigma(\Pg\Pg\to\PA)\mathcal{B}(\PA\to\PZ\Ph\to\ell\ell\PGt\PGt)$, for the $\Pg\Pg\to\PA\to\PZ\Ph$ process.
The model-independent 95\% \CL limits are shown in Fig.~\ref{fig:limits_model_indep} and are consistent with the observed lack of signal.

Model-dependent interpretation of the results is performed in two MSSM scenarios, $\mathrm{M^{125}_{\Ph,EFT}}$ and hMSSM, setting 95\% \CL limits in the $\mA$--$\tanb$ plane.
For both MSSM scenarios, limits are set based on the $\Pg\Pg\to\PA$ and $\bbbar\PA$ production processes.
The signal samples used in the analysis are generated with the $\Pg\Pg\to\PA$ process.
To account for the $\bbbar\PA$ production, at each point in the $\mA$--$\tanb$ plane, the yield of the signal process resulting from $\Pg\Pg\to\PA$
is scaled as follows:
\begin{linenomath}
\begin{equation}
\textrm{Total signal yield} = \Pg\Pg\to\PA\ \text{yield} \left(1 + \epsilon_{\bbbar\PA / \Pg\Pg\to\PA} \frac{\sigma_{\bbbar\PA}}{\sigma_{\Pg\Pg\to\PA}}\right).
\label{eq:scaling_eqn}
\end{equation}
\end{linenomath}
The scaling takes the estimated $\Pg\Pg\to\PA\ \text{yield}$ at each $\mA$--$\tanb$ point and adds a contribution associated to $\bbbar\PA$
according to the estimated selection efficiency ratio in the signal region, $\epsilon_{\bbbar\PA / \Pg\Pg\to\PA} = 0.76$, and
the ratio $\sigma_{\bbbar\PA}/\sigma_{\Pg\Pg\to\PA}$, which depends on $\mA$ and $\tanb$. The signal region selection efficiency ratio was estimated for a single mass point ($\mA = 300\GeV$),
and additional studies were performed to confirm that for the studied mass range (220--400\GeV) the efficiency ratio is nearly flat.
The signal yield scaling allows the estimated $\bbbar\PA$ contribution to be included which is
necessary when setting model-dependent limits in the parameter space region where the $\bbbar\PA$ cross section
becomes nonnegligible compared to the $\Pg\Pg\to\PA$ cross section. For reference, at $\mA = 300\GeV$ and
$\tanb = 4$, in the hMSSM scenario, $\sigma_{\bbbar\PA}/\sigma_{\Pg\Pg\to\PA} = 0.22$, which is a nonnegligible
contribution.

\begin{figure}[h!]
\centering
  \includegraphics[width=0.49\textwidth]{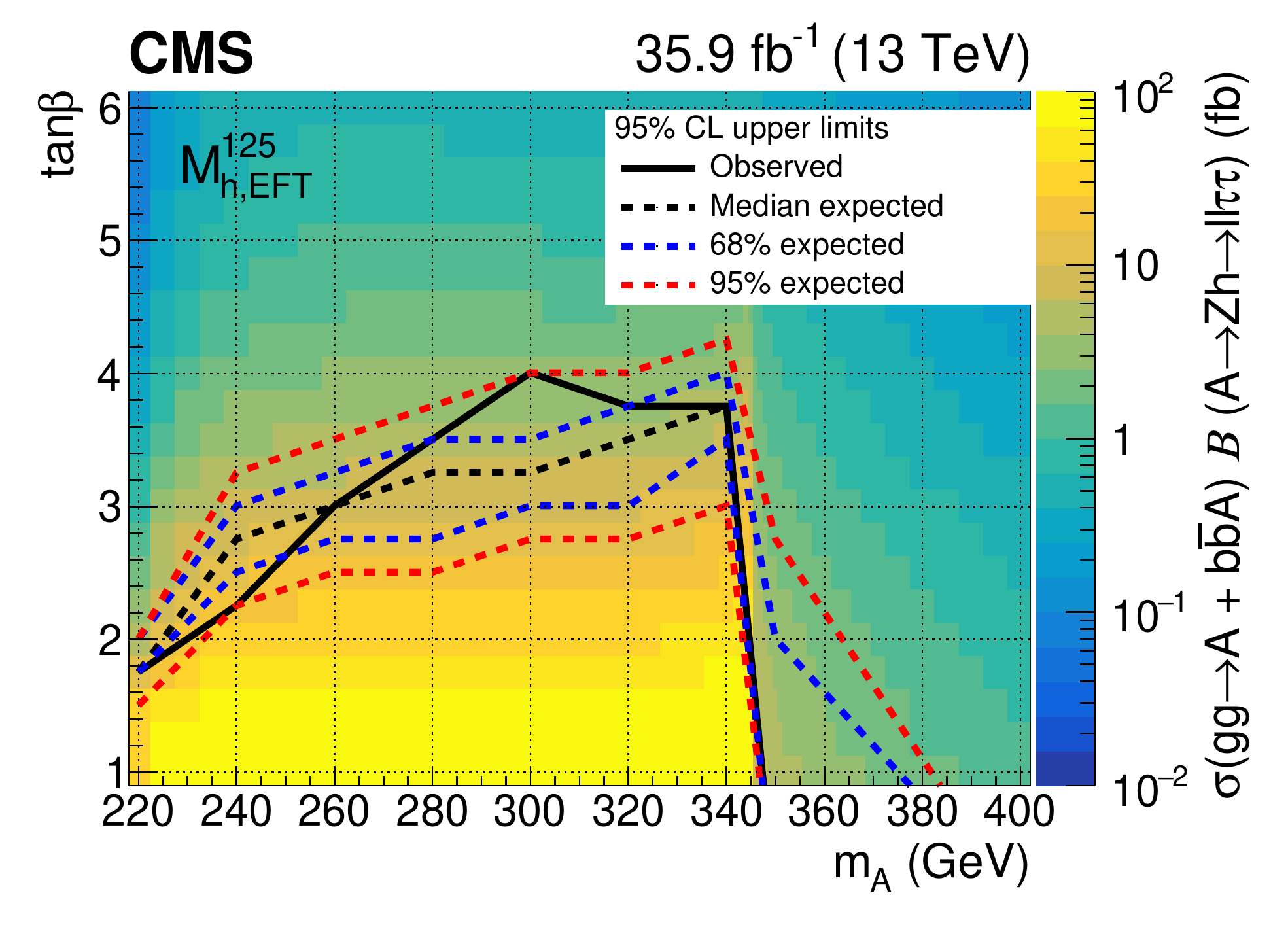}
  \includegraphics[width=0.49\textwidth]{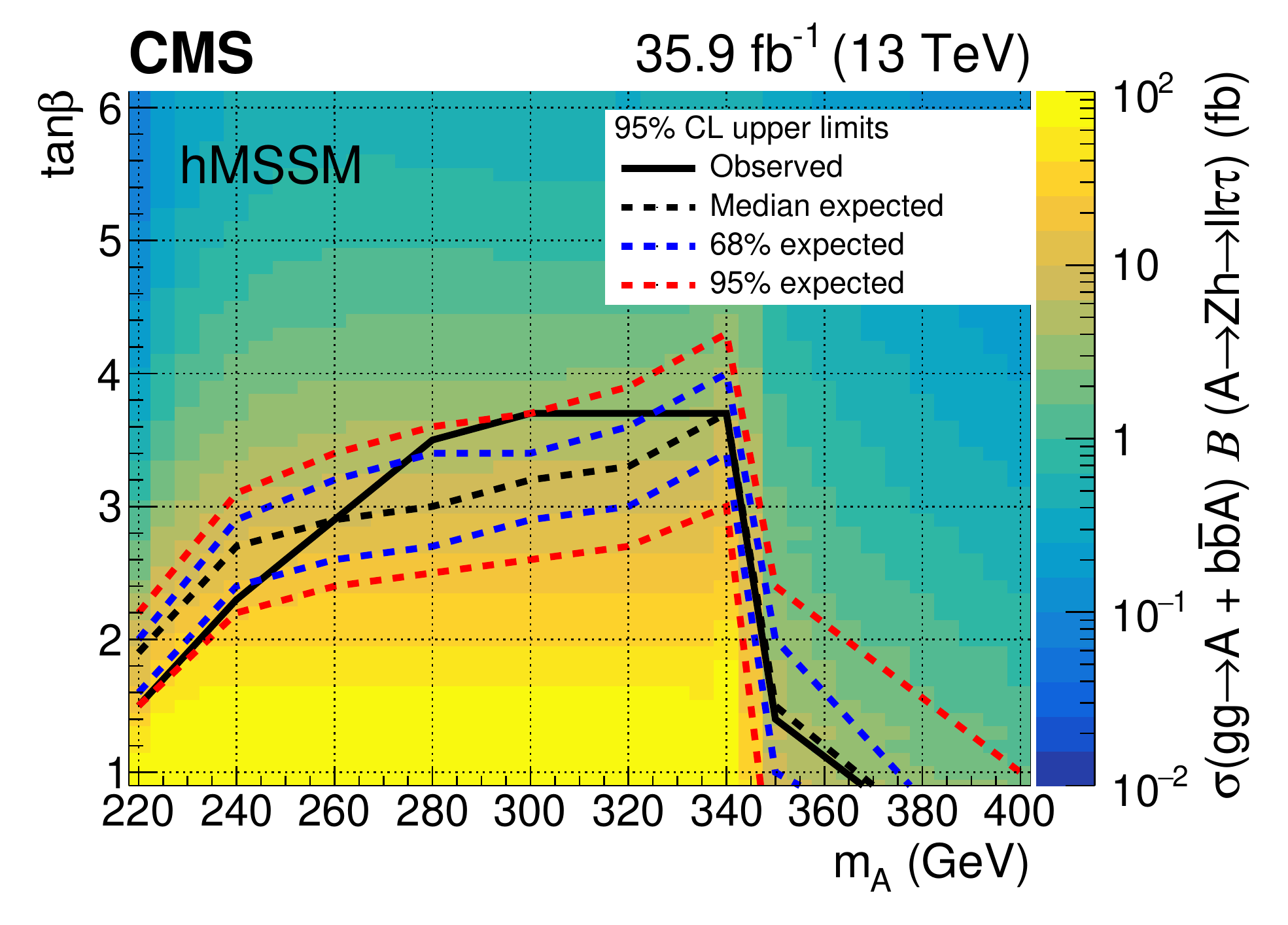}
 \caption{
  The expected and observed 95\% \CL exclusion limits in the $\mA$--$\tanb$ plane
  are shown for two MSSM scenarios:
  $\mathrm{M^{125}_{\Ph,EFT}}$ (left) and hMSSM (right).
  The area under the solid black curve is excluded.
  The dashed black curve corresponds to the median expected limit, surrounded by the 68 (95)\% confidence intervals in blue (red).
  The limits are overlaid on a background showing the
  $\sigma(\Pg\Pg\to\PA+\bbbar\PA)\mathcal{B}(\PA\to\PZ\Ph\to\ell\ell\PGt\PGt)$
  as predicted by each model at each grid point.
 }
 \label{fig:limits_mssm_scenarios}
\end{figure}

The results in the $\mathrm{M^{125}_{\Ph,EFT}}$ scenario and the hMSSM scenario are shown in Fig.~\ref{fig:limits_mssm_scenarios}.
The observed limits exclude slightly higher \tanb values in the $\mathrm{M^{125}_{\Ph,EFT}}$ scenario compared to the hMSSM scenario:
for example at $\mA = 300\GeV$, \tanb values below 4.0 and 3.7 are excluded at 95\% confidence level in the $\mathrm{M^{125}_{\Ph,EFT}}$ and hMSSM scenarios, respectively.

In the hMSSM scenario, this search constrains the parameter space region with low $\tanb$ values when $220<\mA<350\GeV$, and supports the results of previous indirect and direct searches.
The combined measurements of the standard model Higgs boson couplings result in indirect constrains on the hMSSM scenario,
that indicate that $\mA$ values below 600\GeV are disfavored by the observed data~\cite{Aad:2019mbh, Sirunyan:2018koj}.
Out of the direct searches targeting the $\mA$ values below 400\GeV,
this analysis has a similar sensitivity as the searches using the $\PA\to\PZ\Ph (\Ph\to\bbbar)$ decay, performed by the ATLAS and CMS Collaborations~\cite{Aaboud:2017cxo, Sirunyan:2019xls}.
Moreover, together with the results presented in Refs.~\cite{Aaboud:2017cxo, Sirunyan:2019xls}, this analysis complements
the constraints placed by analyses which target the decay of the \PH boson into a pair of \PW or \PZ bosons~\cite{Aaboud:2017gsl, Aaboud:2017rel}.

\section{Summary}
A search is presented for a pseudoscalar Higgs boson decaying into a 125\GeV Higgs boson, which further decays into tau leptons, and a \PZ boson that decays into a pair of electrons or muons.
A data sample of proton-proton collisions collected at $\sqrt{s} = 13\TeV$ by the CMS experiment at the LHC is used, corresponding to an integrated luminosity of 35.9\fbinv.
The sensitivity of the study is increased by using the information on the Higgs boson mass~\cite{Sirunyan:2017exp}
when reconstructing the mass of the pseudoscalar Higgs boson. The signal extraction is further optimized with kinematic selections based on the mass of the Higgs boson.
The data agree with the background predictions from the standard model.
The observed model-independent limits at 95\% confidence level on the product $\sigma(\Pg\Pg\to\PA)\mathcal{B}(\PA\to\PZ\Ph\to\ell\ell\PGt\PGt)$ range
from 27 to $5\unit{fb}$ for $\PA$ boson mass 220 to 400\GeV, respectively.
The model-independent limits are interpreted in terms of $\sigma(\Pg\Pg\to\PA+\bbbar\PA)\mathcal{B}(\PA\to\PZ\Ph\to\ell\ell\PGt\PGt)$ for calculation of the model-dependent limits
in two minimal supersymmetric standard model scenarios, $\mathrm{M^{125}_{\Ph,EFT}}$ and hMSSM.
In the $\mathrm{M^{125}_{\Ph,EFT}}$ (hMSSM) scenario,
the observed limits exclude $\tanb$ values below 1.8 (1.6) at $\mA = 220\GeV$ and 4.0 (3.7) at $\mA = 300\GeV$ at 95\% confidence level.

\begin{acknowledgments}
We congratulate our colleagues in the CERN accelerator departments for the excellent performance of the LHC and thank the technical and administrative staffs at CERN and at other CMS institutes for their contributions to the success of the CMS effort. In addition, we gratefully acknowledge the computing centers and personnel of the Worldwide LHC Computing Grid for delivering so effectively the computing infrastructure essential to our analyses. Finally, we acknowledge the enduring support for the construction and operation of the LHC and the CMS detector provided by the following funding agencies: BMBWF and FWF (Austria); FNRS and FWO (Belgium); CNPq, CAPES, FAPERJ, FAPERGS, and FAPESP (Brazil); MES (Bulgaria); CERN; CAS, MoST, and NSFC (China); COLCIENCIAS (Colombia); MSES and CSF (Croatia); RPF (Cyprus); SENESCYT (Ecuador); MoER, ERC IUT, PUT and ERDF (Estonia); Academy of Finland, MEC, and HIP (Finland); CEA and CNRS/IN2P3 (France); BMBF, DFG, and HGF (Germany); GSRT (Greece); NKFIA (Hungary); DAE and DST (India); IPM (Iran); SFI (Ireland); INFN (Italy); MSIP and NRF (Republic of Korea); MES (Latvia); LAS (Lithuania); MOE and UM (Malaysia); BUAP, CINVESTAV, CONACYT, LNS, SEP, and UASLP-FAI (Mexico); MOS (Montenegro); MBIE (New Zealand); PAEC (Pakistan); MSHE and NSC (Poland); FCT (Portugal); JINR (Dubna); MON, RosAtom, RAS, RFBR, and NRC KI (Russia); MESTD (Serbia); SEIDI, CPAN, PCTI, and FEDER (Spain); MOSTR (Sri Lanka); Swiss Funding Agencies (Switzerland); MST (Taipei); ThEPCenter, IPST, STAR, and NSTDA (Thailand); TUBITAK and TAEK (Turkey); NASU (Ukraine); STFC (United Kingdom); DOE and NSF (USA).

\hyphenation{Rachada-pisek} Individuals have received support from the Marie-Curie program and the European Research Council and Horizon 2020 Grant, contract Nos.\ 675440, 752730, and 765710 (European Union); the Leventis Foundation; the A.P.\ Sloan Foundation; the Alexander von Humboldt Foundation; the Belgian Federal Science Policy Office; the Fonds pour la Formation \`a la Recherche dans l'Industrie et dans l'Agriculture (FRIA-Belgium); the Agentschap voor Innovatie door Wetenschap en Technologie (IWT-Belgium); the F.R.S.-FNRS and FWO (Belgium) under the ``Excellence of Science -- EOS" -- be.h project n.\ 30820817; the Beijing Municipal Science \& Technology Commission, No. Z181100004218003; the Ministry of Education, Youth and Sports (MEYS) of the Czech Republic; the Lend\"ulet (``Momentum") Program and the J\'anos Bolyai Research Scholarship of the Hungarian Academy of Sciences, the New National Excellence Program \'UNKP, the NKFIA research grants 123842, 123959, 124845, 124850, 125105, 128713, 128786, and 129058 (Hungary); the Council of Science and Industrial Research, India; the HOMING PLUS program of the Foundation for Polish Science, cofinanced from European Union, Regional Development Fund, the Mobility Plus program of the Ministry of Science and Higher Education, the National Science Center (Poland), contracts Harmonia 2014/14/M/ST2/00428, Opus 2014/13/B/ST2/02543, 2014/15/B/ST2/03998, and 2015/19/B/ST2/02861, Sonata-bis 2012/07/E/ST2/01406; the National Priorities Research Program by Qatar National Research Fund; the Ministry of Science and Education, grant no. 3.2989.2017 (Russia); the Programa Estatal de Fomento de la Investigaci{\'o}n Cient{\'i}fica y T{\'e}cnica de Excelencia Mar\'{\i}a de Maeztu, grant MDM-2015-0509 and the Programa Severo Ochoa del Principado de Asturias; the Thalis and Aristeia programs cofinanced by EU-ESF and the Greek NSRF; the Rachadapisek Sompot Fund for Postdoctoral Fellowship, Chulalongkorn University and the Chulalongkorn Academic into Its 2nd Century Project Advancement Project (Thailand); the Nvidia Corporation; the Welch Foundation, contract C-1845; and the Weston Havens Foundation (USA).
\end{acknowledgments}

\clearpage

\bibliography{auto_generated}
\cleardoublepage \appendix\section{The CMS Collaboration \label{app:collab}}\begin{sloppypar}\hyphenpenalty=5000\widowpenalty=500\clubpenalty=5000\vskip\cmsinstskip
\textbf{Yerevan Physics Institute, Yerevan, Armenia}\\*[0pt]
A.M.~Sirunyan$^{\textrm{\dag}}$, A.~Tumasyan
\vskip\cmsinstskip
\textbf{Institut f\"{u}r Hochenergiephysik, Wien, Austria}\\*[0pt]
W.~Adam, F.~Ambrogi, T.~Bergauer, J.~Brandstetter, M.~Dragicevic, J.~Er\"{o}, A.~Escalante~Del~Valle, M.~Flechl, R.~Fr\"{u}hwirth\cmsAuthorMark{1}, M.~Jeitler\cmsAuthorMark{1}, N.~Krammer, I.~Kr\"{a}tschmer, D.~Liko, T.~Madlener, I.~Mikulec, N.~Rad, J.~Schieck\cmsAuthorMark{1}, R.~Sch\"{o}fbeck, M.~Spanring, D.~Spitzbart, W.~Waltenberger, C.-E.~Wulz\cmsAuthorMark{1}, M.~Zarucki
% ...
\vskip\cmsinstskip
\dag: Deceased\\
1:  Also at Vienna University of Technology, Vienna, Austria\\
% ...
\end{sloppypar}
\end{document}